\documentclass[12pt]{article}
\usepackage{epsfig}
\newcommand{\beq}{\begin{equation}}
\newcommand{\eeq}{\end{equation}}
\newcommand{\beqa}{\begin{eqnarray}}
\newcommand{\eeqa}{\end{eqnarray}}
\newcommand{\beqar}{\begin{eqnarray*}}
\newcommand{\eeqar}{\end{eqnarray*}}

\begin{document}
\addtolength{\baselineskip}{1.5mm}

\thispagestyle{empty}

\hfill{hep-th/0208068}

\vspace{32pt}

\begin{center}
\textbf{\Large\bf Fluxbranes Delaying Brane-Antibrane Annihilation}

\vspace{48pt}

Hongsu Kim\footnote{hongsu@hepth.hanyang.ac.kr}

\vspace{12pt}

\textit{Department of Physics\\ Hanyang University, Seoul,
133-791, KOREA}

\end{center}
\vspace{48pt}

\begin{abstract}
By intersecting the $RR$ charged $D_{p}-\bar{D}_{p}$ pair $(p=6, 4, 2, 0)$ with the $RR$ $F7$-brane 
and by intersecting the $NSNS$ charged $F1-\bar{F1}$ and $NS5-\bar{NS5}$ pairs with the $NSNS$ 
$F6$-branes, the possibility of stabilizing the brane-antibrane systems is considered.
The behavior of the corresponding supergravity solutions indicates that the $RR$ $F7$-brane 
content of the solution plays the role of keeping the brane and the antibrane from annihilating 
each other completely since the two-brane configuration structure still persists in the vanishing 
inter-brane distance limit of the supergravity solution.
In terms of the stringy description, we interpret this as representing
that the $RR$ $F7$-brane ``delays'' the brane-antibrane annihilation process but only until this 
non-supersymmetric and hence unstable $F7$-brane itself decays. Then next, the behavior of the supergravity 
solutions representing $F1-\bar{F1}$ and $NS5-\bar{NS5}$ again for vanishing inter-brane separation 
reveals that as they approach, these ``$NS$''-charged brane and antibrane always collide and annihilate 
irrespective of the presence or the absence of the $NSNS$ $F6$-brane. And we have essentially
attributed this to the absence of (open) stringy description of the instability in the ``$NS$''-charged case. 
This interpretation may provide a resolution to the contrasting features between the instability of 
``$R$''-charged brane-antibrane systems and that of `$NS$''-charged ones. 
Certainly, however, it poses another puzzle that in the ``$NS$''-charged case, the quantum entity, 
that should take over the semi-classical instability as the inter-brane distance gets smaller, 
is missing. This is rather an embarrassing state of affair that needs to be treated with great care.
\end{abstract}

\hfill{PACS numbers : 11.25.Sq, 04.65.+e, 11.10.Kk}

\setcounter{footnote}{0}

\newpage

\section{Introduction}

In the present work, we would like to address the issue of possible flux stabilization of the unstable, 
non-BPS $D_{p}-\bar{D}_{p}$ system \cite{non-BPS}. Thus it might be relevant to first remind why the $D_{p}-\bar{D}_{p}$ 
system is unstable to begin with by resorting to a simple argument that
goes as follows. Consider a system consisting of a certain number $N$ of coincident $D_{P}$-branes separated by some distance from a
system of $N$ coincident $\bar{D}_{p}$-branes, for simplicity, in flat $R^{10}$. This system differs from the BPS system of $2N$
$D_{p}$-branes by the orientation reversal on the antibranes. In this system, the branes and the antibranes each break a different
half of the original supersymmetry and the whole configuration is non-supersymmetric or non-BPS and hence is unstable. As a result,
there is a combined gravitational and (RR) gauge attractive force between the branes and the antibranes at some large but finite
separation leading to the semi-classical instability. At the separation of order the string scale, $\sim \sqrt{\alpha'}=l_{s}$,
in particular, the open string connecting a $D_{p}$-brane to a $\bar{D}_{p}$-brane becomes tachyonic. What then would be the
eventual fate or endpoint of this unstable $D_{p}-\bar{D}_{p}$-system ? According to Sen \cite{sen1}, the endpoint could be the
supersymmetric vacuum via the open string tachyon condensation. To be a little more concrete, in the $D_{p}-\bar{D}_{p}$ annihilation
process, the open string tachyon behaves as a Higgs field and condenses to a minimum of its potential, breaking the worldvolume
gauge symmetry to its diagonal subgroup,
\beq
U(N) \times U(N) \rightarrow U(N). \nonumber
\eeq
Now, if the outcome of this brane-antibrane annihilation were, as advocated by Sen \cite{sen1}, the supersymmetric vacuum, the residual
gauge symmetry, i.e., the diagonal subgroup $U(N)$, should also disappear, presumably by the process suggested by Sen \cite{sen1} 
or by Yi \cite{yi}. Regarding the
conjectures on the possible endpoints of the unstable $D_{p}-\bar{D}_{p}$-systems, it is interesting to note that there are some
suggestions on the obstructions against complete annihilation of the $D_{p}-\bar{D}_{p}$ system. One is the argument that due to the
topological difference between the Chan-Paton bundle $E$ carried by the $D_{p}$-branes and $F$ carried by $\bar{D}_{p}$-branes, the
endpoint could be a lower-dimensional D-brane instead. The other is the suggestion that endpoint could be a stable $D$-brane as a
topological defect (soliton) arising in the worldvolume Higgs mechanism (i.e., gauge symmetry breaking) $U(N)\times U(N) \rightarrow
U(N)$, namely the tachyon condensation, classified by the homotopy group
\beq
\Pi_{2n-1}(U(N)) = Z,  ~~~\Pi_{2n}(U(N)) = 0, \nonumber
\eeq
with $N$ in the stable regime \cite{horava}. Having been convinced of the generic instability of the $D_{p}-\bar{D}_{p}$ system which
is quantum (in terms of tachyon condensation) for vey small brane-antibrane separation and is semi-classical in nature for large but
finite separation, in the present work, we would like to discuss the possibility of stabilizing the brane-antibrane systems by
intersecting the $RR$ charged $D_{p}-\bar{D}_{p}$ pair $(p=6, 4, 2, 0)$ with the $RR$ $F7$-brane \cite{costa, flux, russo1} 
(which will be denoted henceforth by $(D_{p}-\bar{D}_{p})||F7$) and intersecting $NSNS$ charged $F1-\bar{F1}$ and $NS5-\bar{NS5}$ 
pairs  with the $NSNS$ $F6$-branes \cite{russo1} (which similarly will be denoted by $(F1-\bar{F1})||F6$ and 
$(NS5-\bar{NS5})||F6$ respectively). Since it is the $D6$ (or $\bar{D6}$) brane which has
non-trivial coupling to the magnetic flux of $RR$ $F7$-brane and the $NS5$ (or $\bar{NS5}$) brane that couples directly to that of the
$NSNS$ $F6$-brane, one might naturally expect that only the $D6-\bar{D6}$ and $NS5-\bar{NS5}$ systems, but no others, would be
balanced, when the $RR$ $F7$-brane and the $NSNS$ $F6$-brane are intersected respectively, in an (unstable) equilibrium against
the combined gravitational and gauge attractions. Within the context of analysis based on the explicit supergravity solutions,
however, things turn out not to be so transparent. As we shall see in a moment, although the $RR$ $F7$-brane
fails to serve to stabilize the $D_{p}-\bar{D}_{p}$ pairs $(p=4, 2, 0)$ against the collapse
for lage but finite separation, when the branes and the antibranes are brought close enough together, the fluxbranes play the role
of keeping them from merging and then annihilating each other completely. And the resolution to this apparent puzzle lies in the
validity of the semi-classical supergravity description of the system. Namely, as the inter-brane separation gets smaller and smaller,
say, towards the order of string scale $\sim \sqrt{\alpha'}=l_{s}$, the supergravity description becomes no longer reliable and the
semi-classical instability should be replaced by the quantum, stringy instability expressed in terms of the open and closed string
tachyon condensation. The conclusion we shall draw is that, if stated briefly, the $RR$ $F7$-brane simply {\it delays} the
annihilation process of the $D_{p}-\bar{D}_{p}$ systems only until the unstable $F7$-brane itself decays toward either a supersymmetric
string vacuum or the nucleation of the $D6-\bar{D6}$ pairs via the brany Schwinger process \cite{costa}. 
And here, we noticed the facts that
firstly, the $F7$-brane breaks all the supersymmetries and hence should be unstable and decay \cite{costa, flux} and secondly, $D_{(p-1)}-\bar{D}_{(p-1)}$
pairs can generally be created from the $RR$ $F_{p}$-brane background via the brane-version of Schwinger process. And eventually, 
the $RR$ fluxbrane can never eliminate the instability of the brane-antibrane system completely and hence the endpoint
of these $(p|(D_{p}-\bar{D}_{p}), F7)$ systems $(p=4, 2, 0)$ would be either the supersymmetric vacuum or lower-dimensional branes
arising as a result of topological obstruction argument given earlier. The $D6-\bar{D6}$ system which stands out as a unique case,
however, would be supported by the $F7$-brane against collapse. But again, this would be true only within the time scale for the decay
of $F7$-brane and once the $F7$-brane decays presumably leaving a supersymmetric closed string vacuum behind, even $D6$ and $\bar{D6}$
would collide and annihilate each other leaving yet another supersymmetric vacuum behind.

\section{$D-anti-D$ systems supported by $RR$ fluxbrane}\label{ }

In this section, we shall consider the intersection of non-BPS $D_{2p}-\bar{D}_{2p}$ systems with 
the magnetic $RR$ $F7$-brane in IIA theory in order to study the role played by the $RR$ $F7$-brane
concerning the semi-classical and quantum (in terms of open string tachyon condensation)
instability of the $D_{2p}-\bar{D}_{2p}$ systems.
We first begin with the exact supergravity solutions
representing the $D6-\bar{D6}$ system and
the intersection of this $D6-\bar{D6}$ with a magnetic $RR$ flux
7-brane ($(D6-\bar{D6})||F7$ for short). The former
possesses conical singularities which can be made to disappear by
introducing the $RR$ magnetic field (i.e., the $RR$ $F7$-brane)
and properly tuning its strength in the latter. In order to
demonstrate this, we need to along the way perform the M-theory
uplift of the $D6-\bar{D6}$ pair which leads to the Kaluza-Klein
($KK$) monopole/anti-monopole solution ($KK-dipole$ henceforth)
first discussed in the literature by Sen \cite{sen2} by embedding
the Gibbons-Perry \cite{gp} $KK-dipole$ solution in $D=11$
M-theory context. Indeed the $D6$-brane solution is unique among
$D_{p}$-brane solutions in IIA/IIB theories in that it is a
codimension 3 object and hence in many respects behaves like the
familiar abelian magnetic monopole in $D=4$. This, in turn,
implies that the $D6-\bar{D6}$ solution should exhibit essentially
the same generic features as those of Bonnor's magnetic dipole
solution \cite {bonnor1, emp} and its dilatonic generalizations \cite{bonnor2, emp}
in $D=4$ studied extensively in the recent literature. As we shall
see in a moment, these similarities allow us to envisage the
generic nature of instabilities common in all unstable
$D_{p}-\bar{D}_{p}$ systems in a simple and familiar manner.
Then next, we consider the exact supergravity solutions
representing the electrically $RR$-charged $D0-\bar{D0}$ system and
the intersection of this $D0-\bar{D0}$ with a magnetic $RR$ flux
7-brane. This last system as well as $(D2-\bar{D2})||F7$ and $(D4-\bar{D4})||F7$
systems exhibit rather puzzling features and we attempt to provide resolutions
to them later on.

\subsection{$D6-\bar{D6}$ pair supported by $RR$ $F7$-brane}

{\bf \large (A) $D6-\bar{D6}$ pair in the absence of the magnetic field}

In string frame, the exact IIA supergravity solution representing
the $D6-\bar{D6}$ pair is given by \cite{youm, emp} 
\beqa
ds^2_{10} &=& H^{-1/2}[-dt^2 + \sum^{6}_{i=1} dx^2_{i}] +
H^{1/2}[(\Delta+a^2\sin^2 \theta)\left({dr^2\over
\Delta}+d\theta^2\right) + \Delta \sin^2 \theta d\phi^2],
\nonumber \\ 
e^{2\phi} &=& H^{-3/2}, \\ 
A_{[1]} &=& \left[{2mra\sin^2 \theta \over {\Delta + a^2\sin^2
\theta}}\right]d\phi, ~~~F_{[2]}=dA_{[1]} \nonumber 
\eeqa 
where the harmonic function in 3-dimensional transverse space is given
by 
\beq 
H(r) = {\Sigma \over {\Delta + a^2\sin^2 \theta}} 
\eeq 
and $\Sigma = r^2-a^2\cos^2 \theta$, $\Delta = r^2-2mr-a^2$ with $r$
being the radial coordinate in the transverse directions. The
parameter $a$ can be thought of as representing the separation
between the brane and antibrane (we will elaborate on this
shortly) and changing the sign of $a$ amounts to reversing the
orientation of the brane pair, so here we will choose, without
loss of generality, $a\geq 0$. $m$ is the ADM mass of each brane
and the ADM mass of the whole $D6-\bar{D6}$ system is $M_{ADM}=2m$
which should be obvious as it would be the sum of ADM mass of each
brane when they are well separated. It is also noteworthy that,
similarly to what happens in the Ernst solution \cite{ernst} in
$D=4$ Einstein-Maxwell theory describing a pair of
oppositely-charged black holes accelerating away from each other
(due to the Melvin magnetic universe content), this $D6-\bar{D6}$
solution in IIA theory is also static but {\it axisymmetric} in
these Boyer-Lindquist-type coordinates. As has been pointed out by
Sen \cite{sen2} in the M-theory $KK-dipole$ solution case and by
Emparan \cite{emp} in the case of generalized Bonnor's solution,
the IIA theory $D6-\bar{D6}$ solution given above represents the
configuration in which a $D6$-brane and a $\bar{D6}$-brane are
sitting on the endpoints of the dipole, i.e., $(r=r_{+},
\theta=0)$ and $(r=r_{+}, \theta=\pi)$ respectively where $r_{+}$
is the larger root of $\Delta =0$, namely
$r_{+}=m+\sqrt{m^2+a^2}$. Next, we turn to the conical singularity
structure of this $D6-\bar{D6}$ solution. First observe that the
rotational Killing field $\psi^{\mu} = (\partial/\partial
\phi)^{\mu}$ possesses vanishing norm, i.e., $\psi^{\mu}\psi_{\mu}
= g_{\alpha \beta}\psi^{\alpha}\psi^{\beta} = g_{\phi\phi} = 0$ at
the locus of $r=r_{+}$ as well as along the semi-infinite lines
$\theta=0, \pi$. This implies that $r=r_{+}$ can be thought of as
a part of the symmetry axis of the solution. Namely unlike the
other familiar axisymmetric solutions, for the case of the
$D6-\bar{D6}$ solution under consideration, the endpoints of the
two semi-axes $\theta =0$ and $\theta =\pi$ do not come to join at
a common point. Instead, the axis of symmetry is completed by the
segment $r=r_{+}$. And as $\theta$ varies from $0$ to $\pi$, one
moves along the segment from $(r=r_{+}, \theta=0)$ where $D6$ is
situated to $(r=r_{+}, \theta=\pi)$ where $\bar{D6}$ is placed.
Then the natural question to be addressed is whether or not the
conical singularities arise on different portions of the symmetry
axis. This situation is very reminiscent of the conical
singularity structure in the generalized Bonnor's dipole solution
in Einstein-Maxwell-dilaton theory in $D=4$ extensively studied by
Emparan \cite{emp} recently. Thus below, we explore the nature of
possible conical singularities in this $D6-\bar{D6}$ solution in
$D=10$ type IIA theory following essentially the same avenue as
that presented in the work of Emparan \cite{emp}. Namely, consider
that ; if $C$ is the proper length of the circumference around the
symmetry axis and $R$ is its proper radius, then the occurrence of
a conical angle deficit (or excess) $\delta$ would manifest itself
if $(dC/dR)|_{R\rightarrow 0}=2\pi - \delta$. We now proceed to
evaluate this conical deficit (or excess) assuming first that the
azimuthal angle coordinate $\phi$ is identified with period
$\Delta \phi$. The conical deficit along the axes $\theta =0, \pi$
and along the segment $r=r_{+}$ are given respectively by 
\beqa
\delta_{(0,\pi)} &=& 2\pi - \arrowvert {\Delta \phi
d\sqrt{g_{\phi\phi}} \over \sqrt{g_{\theta\theta}}d\theta}
\arrowvert_{\theta =0, \pi} = 2\pi - \Delta \phi, \\
\delta_{(r=r_{+})} &=& 2\pi - \arrowvert {\Delta \phi
d\sqrt{g_{\phi\phi}} \over \sqrt{g_{rr}}dr} \arrowvert_{r=r_{+}} =
2\pi - \left(1 + {m^2\over a^2}\right)^{1/2}\Delta \phi \nonumber
\eeqa 
where, of course, we used the $D6-\bar{D6}$ metric solution
given in eq.(3). From eq.(5), it is now evident that one cannot
eliminate the conical singularities along the semi-axes $\theta =
0, \pi$ and along the segment $r=r_{+}$ at the same time. Indeed
one has the options : \\ 
(i) One can remove the conical angle
deficit along $\theta = 0, \pi$ by choosing $\Delta \phi = 2\pi$
at the expense of the conical angle excess along $r=r_{+}$ which
amounts to the presence of a {\it strut} providing the internal
pressure to counterbalance the combined gravitational and gauge
attractions between $D6$ and $\bar{D6}$. \\ 
(ii) Alternatively,
one can instead eliminate the conical singulatity along $r=r_{+}$
by choosing $\Delta \phi = 2\pi(1+m^2/a^2)^{-1/2}$ at the expense
of the appearance of the conical angle deficit
$\delta_{(0,\pi)}=2\pi[1-\{a^2/(m^2+a^2)\}^{1/2}]$ along $\theta
=0, \pi$ which implies the presence of {\it cosmic strings}
providing the tension \beq \tau ={\delta_{(0,\pi)}\over 8\pi} =
{1\over 4}\left[1 - \left({a^2\over m^2+a^2}\right)^{1/2}\right]
\nonumber \eeq that pulls $D6$ and $\bar{D6}$ at the endpoints
apart. \\ Normally, one might wish to take the second option in
which the pair of branes is suspended by open cosmic strings,
namely $D6$ and $\bar{D6}$ are kept apart by the tensions
generated by cosmic strings against the collapse due to the
gravitational and gauge attractions. And the line $r=r_{+}$,
$0<\theta <\pi$ joining $D6$ and $\bar{D6}$ is now completely
non-singular. This recourse to cosmic strings to account for the
conical singularities of the solution and to suspend the
$D6-\bar{D6}$ system in an equilibrium configuration, however,
might appear as a rather {\it ad hoc} prescription. Perhaps it
would be more relevant to introduce an external magnetic field
aligned with the axis joining the brane pair to counterbalance the
combined gravitational and gauge attractions by pulling them
apart. By properly {\it tuning} the strength of the magnetic
field, the attractive inter-brane force along the axis would be
rendered to vanish. Indeed this conical singularity structure of
the $D6-\bar{D6}$ system and its cure via the introduction of the
external magnetic field of proper strength is reminiscent of
Ernst's prescription \cite{ernst} for the elimination of conical
singularities of the charged $C$-metric and of Emparan's treatment
\cite{emp} to remove the analogous conical singularities of the
Bonnor's magnetic dipole solution in Einstein-Maxwell and
Einstein-Maxwell-dilaton theories and in the present work, we
shall closely follow the formulation of Emparan \cite{emp}. Before
doing so, however, we need to study the geometrical structure of
the $D6-\bar{D6}$ system in IIA theory given in eqs.(3) and (4) in
some more detail.

(1) {\it \large The meaning of parameter $a$ as the proper inter-brane
distance}

We now elaborate on our earlier comment that the parameter $a$
appearing in this supergravity solution can be regarded as
indicating the proper separation between the brane and the
antibrane. Notice first that for large $a$, the proper inter-brane
distance increases as $\sim 2a$. Namely, 
\beqa 
l &=&
\int^{\pi}_{0}d\theta \sqrt{g_{\theta\theta}}|_{r=r_{+}} =
\int^{\pi}_{0}d\theta H^{1/4}(\Delta + a^2\sin^2
\theta)^{1/2}|_{r=r_{+}} \nonumber \\ &\simeq &
\int^{\pi}_{0}d\theta a\sin \theta = 2a. 
\eeqa 
Meanwhile, as
argued by Sen \cite{sen2}, the proper inter-brane distance
vanishes when $a\rightarrow 0$. In addition, that the limit
$a\rightarrow 0$ actually amounts to the vanishing inter-brane
distance can be made more transparent as follows. Recently, Brax,
Mandal and Oz \cite{bmo} discussed the supergravity solution
representing {\it coincident} $D_{p}-\bar{D}_{p}$ pairs in type II
theories and studied its instability in terms of the condensation
of tachyon arising in the spectrum of open strings stretched
between $D_{p}$ and $\bar{D}_{p}$. Thus now, taking the $p=6$ case
for example, we first would like to establish the correspondence
between our solution given above representing $D6-\bar{D6}$ pair
generally separated by an arbitrary distance and theirs. For
specific but appropriate values of the parameters appearing in
their solution, $(c_{2}=1 (p>3), c_{1}=0, r_{0}=m/2)$ \cite{bmo}
so as to represent a neutral, coincident $D6-\bar{D6}$ pair, their
solution is given in Einstein frame by
\beqa
ds^2_{E} &=& \left[{1-r_{0}/\tilde{r} \over
1+r_{0}/\tilde{r}}\right]^{1/4}[-dt^2 + \sum^{6}_{i=1} dx^2_{i}] +
\left[1-{r_{0}\over \tilde{r}}\right]^{1/4}\left[1+{r_{0}\over
\tilde{r}}\right]^{15/4} [d\tilde{r}^2 + \tilde{r}^2(d\theta^2 +
\sin^2 \theta d\phi^2)], \nonumber \\ e^{\phi} &=&
\left[{1-r_{0}/\tilde{r} \over 1+r_{0}/\tilde{r}}\right]^{3/2},
~~~A_{[1]} = 0 \eeqa which, in string frame, using
$g^{E}_{\mu\nu}=e^{-\phi/2}g^{S}_{\mu\nu}$, becomes \beqa ds^2 &=&
\left[{1-r_{0}/\tilde{r} \over 1+r_{0}/\tilde{r}}\right][-dt^2 +
\sum^{6}_{i=1} dx^2_{i}] + \left[1-{r_{0}\over
\tilde{r}}\right]\left[1+{r_{0}\over \tilde{r}}\right]^{3}
[d\tilde{r}^2 + \tilde{r}^2(d\theta^2 + \sin^2 \theta d\phi^2)],
\nonumber \\ e^{\phi} &=& \left[{1-r_{0}/\tilde{r} \over
1+r_{0}/\tilde{r}}\right]^{3/2}, ~~~A_{[1]} = 0. \eeqa Consider
now, transforming from this {\it isotropic} coordinate $\tilde{r}$
to the standard radial $r$ coordinate \beq \tilde{r} = {1\over
2}[(r-2r_{0}) + (r^2-4r_{0}r)^{1/2}] ~~~{\rm or ~inversely}
~~~r=\tilde{r}\left(1+{r_{0}\over \tilde{r}}\right)^2.
\eeq
Then their solution describing $(N=1)D6$ and $(\bar{N}=1)\bar{D6}$ now
takes the form
\beqa
ds^2 &=& \left(1-{2m\over r}\right)^{1/2}[-dt^2 + \sum^{6}_{i=1} dx^2_{i}] +
\left(1-{2m\over r}\right)^{-1/2}[dr^2 + r^2\left(1-{2m\over
r}\right)(d\theta^2 + \sin^2 \theta d\phi^2)], \nonumber \\
e^{2\phi} &=& \left(1-{2m\over r}\right)^{3/2}, ~~~A_{[1]} = 0.
\eeqa
Clearly, this solution indeed coincides with the
$a\rightarrow 0$ (i.e., vanishing separation) limit of our more
general $D6-\bar{D6}$ solution given in eq.(3). Actually, this
aspect also has been pointed out in a recent literature \cite
{teo}. And this confirms our earlier proposition that $a$ indeed
acts as a relevant parameter representing the proper inter-brane
distance even for very small separation.

(2) {\it \large The geometry near each pole of $D6-\bar{D6}$ pair}

Thus far, we have simply accepted that the supergravity solution
given in eqs.(3),(4) represents the configuration of $D6-\bar{D6}$
pair. It would therefore be satisfying to demonstrate in a
transparent manner that this is indeed the case. To this end,
first note that the solution in eq.(3) clearly is
asymptotically-flat as $r\rightarrow \infty$ and in this
asymptotic region, the $RR$ tensor potential is indeed that of a
``dipole'', i.e., $A_{[1]} \rightarrow {2ma\over r}sin^2\theta
d\phi = {ma\over r}(1-\cos\theta)d\phi$. Also note that the axis
of symmetry of the solution (i.e., the fixed point set of the
isometry generated by the Killing field $(\partial/\partial
\phi)$) consists of the semi-infinite lines $\theta = 0, \pi$
(running from $r=r_{+}$ to $r=\infty$) and the segment $r=r_{+}$
(running from $\theta = 0$ to $\theta = \pi$). And indeed at each
of the poles, ($r=r_{+}, \theta=0$) and ($r=r_{+}, \theta=\pi$),
lies a (distorted) brane and (distorted) antibrane respectively.
Thus in order to show this explicitly, we perform the change of
coordinates from ($r, \theta$) to ($\rho, \bar{\theta}$) given by
the following transformation law \cite{sen2, emp} 
\beq 
r = r_{+} +
{\rho \over 2}(1+\cos \bar{\theta}), ~~~\sin^2 \theta = {\rho
\over \sqrt{m^2+a^2}}(1-\cos \bar{\theta}) 
\eeq 
where
$r_{+}=m+\sqrt{m^2+a^2}$ as given earlier. We start with the study
of the metric near $(r=r_{+}, \theta=0)$. Upon changing the
coordinates as given by eq.(12) and then taking $\rho$ to be much
smaller than any other length scale involved so as to get near
each pole, the $D6-\bar{D6}$ solution in eq.(3) becomes 
\beqa
ds^2_{10} &\simeq & g^{1/2}(\bar{\theta})\left({\rho\over
q}\right)^{1/2}[-dt^2 + \sum^{6}_{i=1} dx^2_{i}] + \left({q\over
\rho}\right)^{1/2}[ g^{1/2}(\bar{\theta})(d\rho^2 +
\rho^2d\bar{\theta}^2) + g^{-1/2}(\bar{\theta})\rho^2 \sin^2
\bar{\theta}d\phi^2], \nonumber \\ e^{2\phi} &\simeq &
\left({\rho\over q}\right)^{3/2} g^{3/2}(\bar{\theta}), \\
A^{m}_{[1]} &\simeq & {mr_{+}a\over
(m^2+a^2)}g^{-1}(\bar{\theta})(1-\cos \bar{\theta})d\phi \nonumber
\eeqa 
where $q=mr_{+}/\sqrt{m^2+a^2}$ is the $RR$ charge of each $D6$-brane 
and $g(\bar{\theta}) = cos^2(\bar{\theta}/2) + [a^2/(m^2+a^2)]\sin^2(\bar{\theta}/2)$.
Namely in this small-$\rho$ limit, the geometry of the solution
reduces to that of the near-horizon limit of a $D6$-brane.
However, the horizon is no longer spherically-symmetric and is
deformed due to the presence of the other brane, i.e., the
$\bar{D6}$-brane located at the other pole. To elaborate on this
point, it is noteworthy that the surface $r=r_{+}$ is still a
horizon, but instead of being spherically-symmetric, it is
elongated along the axis joining the poles in a prolate shape.
Namely, the horizon turns out to be a {\it prolate spheroid} with
the distortion factor given by $g(\bar{\theta})$. And of course,
it is further distorted by a conical defect at the poles. And
similar analysis can be carried out near the other pole at which
$\bar{D6}$ is situated, i.e., near $(r=r_{+}, \theta=\pi)$. The
limiting geometries above were valid for arbitrary values of
``$a$'', as long as we remain close to each pole. If instead we
consider the limit of very large-$a$, while keeping $(r-r_{+})$
and $a\sin^2\theta$ finite, the supergravity solution in eq.(3)
reduces, in this time, to 
\beqa 
ds^2_{10} &\simeq & \left(1+{q\over \rho}\right)^{-1/2}[-dt^2 +
\sum^{6}_{i=1}dx^2_{i}] + \left(1+{q\over
\rho}\right)^{1/2}[d\rho^2 + \rho^2 (d\bar{\theta}^2 +
\sin^2\bar{\theta}d\phi^2)], \nonumber \\ 
e^{2\phi} &\simeq & \left(1+{q\over \rho}\right)^{-3/2}, \\ 
A^{m}_{[1]} &\simeq & q(1 - \cos\bar{\theta})d\phi \nonumber 
\eeqa 
with $q\rightarrow m$.
Clearly, this can be recognized as representing the extremal
$D6$-brane solution with $\rho^2 = \sum^{9}_{m=7}x^2_{m}$. Indeed,
this result was rather expected since, physically, taking the
limit $a\rightarrow \infty$ amounts to pushing one of the poles
(say, $\bar{D6}$-brane) to a large distance and studying the
geometry of the remaining pole ($D6$-brane) which, as a
consequence, should be spherically-symmetric. So we conclude that
the solution given in eq.(3) indeed describes a {\it dipole},
i.e., the $D6-\bar{D6}$ pair.

{\bf \large (B) $D6-\bar{D6}$ pair in the presence of the magnetic field}

In order to introduce the external magnetic field with proper
strength to counterbalance the combined gravitational and gauge
attractions and hence to keep the $D6-\bar{D6}$ pair in an
(unstable) equilibrium configuration, we now proceed to construct
the supergravity solution representing $D6-\bar{D6}$ pair
parallely intersecting with a $RR$ $F7$-brane. This can be achieved by
first uplifting the $D6-\bar{D6}$ solution in IIA theory to the
$D=11$ $KK-dipole$ solution in M-theory discussed by Sen
\cite{sen2} and then by performing a {\it twisted} KK-reduction on
this M-theory $KK-dipole$. Thus consider carrying out the
dimensional lift of the $D6-\bar{D6}$ solution given in eqs.(3) and (4) to
$D=11$ via the standard KK-ansatz
\beq
ds^2_{11} = e^{-{2\over 3}\phi}ds^2_{10} + e^{{4\over 3}\phi}(dy+A_{\mu}dx^{\mu})^2
\eeq
(with $A_{[1]}=A_{\mu}d^{\mu}$ being the 1-form magnetic $RR$
potential in eq.(4)) which yields
\beqa
ds^2_{11} &=& [-dt^2 + \sum^{6}_{i=1} dx^2_{i}] +
\Sigma\left[{dr^2\over \Delta} + d\theta^2\right] \\ &+&
{1\over\Sigma}[\Delta (dy-a\sin^2 \theta d\phi)^2 + \sin^2 \theta
\{(r^2-a^2)d\phi + ady\}^2]. \nonumber
\eeqa
This is the $KK$ monopole/anti-monopole solution in $D=11$ or the
M-theory $KK-dipole$ solution first given by Sen \cite{sen2}.
Similarly to the IIA theory $D6-\bar{D6}$ solution discussed
above, this M-theory $KK-dipole$ solution represents the
configuration in which $KK$ monopole and anti-monopole are sitting
on the endpoints of the dipole, i.e., $(r=r_{+}, \theta=0)$ and
$(r=r_{+}, \theta=\pi)$ respectively. Note that unlike the IIA
theory $D6-\bar{D6}$ solution, this M-theory $KK-dipole$ solution
is free of conical singularities provided the azimuthal angle
coordinate $\phi$ is periodically identified with the standard
period of $2\pi$. Now to get back down to $D=10$, consider
performing the non-trivial point identification \cite{flux}
\beq 
(y, ~\phi)\equiv (y + 2\pi n_{1}R, ~\phi + 2\pi n_{1}RB + 2\pi n_{2}) 
\eeq
(with $n_{1}, ~n_{2}\in Z$) on the M-theory $KK-dipole$ solution
in eq.(16), followed by the associated skew KK-reduction along the
orbit of the Killing field \beq l = \left(\partial/\partial
y\right) + B\left(\partial/\partial \phi \right) \eeq where $B$ is
a magnetic field parameter. And this amounts to introducing the
{\it adapted} coordinate \beq \tilde{\phi} = \phi - By \eeq which
is constant along the orbits of $l$ and possesses standard period
of $2\pi$ and then proceeding with the standard KK-reduction along
the orbit of $(\partial/\partial y)$. Thus we recast the metric
solution, upon changing to this adapted coordinate, in the
standard KK-ansatz
\beqa
ds^2_{11} &=& [-dt^2 + \sum^{6}_{i=1}dx^2_{i}] + \Sigma\left[{dr^2\over \Delta} +
d\theta^2\right] + {\Delta + a^2\sin^2 \theta \over \Sigma}dy^2
\nonumber \\
&+& {2[(r^2-a^2)-\Delta]a\sin^2 \theta \over
\Sigma}dy(d\tilde{\phi} + Bdy) + {\sin^2 \theta \over
\Sigma}[(r^2-a^2)^2+\Delta a^2\sin^2 \theta](d\tilde{\phi} +
Bdy)^2 \nonumber \\
&=& e^{-{2\over 3}\phi}ds^2_{10} + e^{{4\over
3}\phi}(dy+A_{\mu}dx^{\mu})^2
\eeqa
and then read off the $10$-dimensional fields as
\beqa
ds^2_{10} &=& \Lambda^{1/2}\left\{[-dt^2 + \sum^{6}_{i=1}dx^2_{i}] +
\Sigma\left[{dr^2\over \Delta} + d\theta^2\right]\right\} +
\Lambda^{-1/2}\Delta \sin^2 \theta d\tilde{\phi}^2, \nonumber \\
e^{{4\over 3}\phi} &=& \Lambda, \\
A_{[1]} &=& \Lambda^{-1}{\sin^2\theta \over \Sigma}\left\{B[(r^2-a^2)^2+\Delta a^2\sin^2 \theta]
+ a[(r^2-a^2)-\Delta]\right\} d\tilde{\phi}, \nonumber \\
F_{[2]} &=& (\partial_{r}A_{\tilde{\phi}})dr\wedge d\tilde{\phi} +
(\partial_{\theta}A_{\tilde{\phi}})d\theta\wedge d\tilde{\phi},
~~~{\rm where} \nonumber \\
\Lambda &=& {1\over \Sigma}\left\{[\Delta+a^2\sin^2\theta]+
2Ba\sin^2\theta[(r^2-a^2)-\Delta]+B^2\sin^2\theta
[(r^2-a^2)^2+\Delta a^2\sin^2\theta]\right\}. \nonumber
\eeqa
Note that this solution can be identified with a $D6-\bar{D6}$ pair
parallely intersecting with a magnetic $RR$ $F7$-brane since for
$B=0$, it reduces to the $D6-\bar{D6}$ solution in eq.(3) while
for $m=0$ and $a=0$, it reduces to a $RR$ $F7$-brane solution in
IIA theory. To see this last point explicitly, we set $m=0=a$ in
eq.(21) to get
\beqa
ds^2_{10} &=& \Lambda^{1/2}\left[-dt^2 + \sum^{6}_{i=1}dx^2_{i} +
dr^2 + r^2d\theta^2\right] + \Lambda^{-1/2}r^2\sin^2\theta^2
d\tilde{\phi}^2, \nonumber \\ 
e^{2\phi} &=& \Lambda^{3/2}, \\
A_{[1]} &=& A_{\tilde{\phi}}d\tilde{\phi} = {Br^2\sin^2 \theta
\over (1+B^2r^2\sin^2 \theta)} ~~~{\rm where ~~now} \nonumber \\
\Lambda &=& (1 + B^2r^2\sin^2 \theta). \nonumber
\eeqa
Clearly, this is a magnetic $RR$ $F7$-brane solution in type IIA
theory. Also note that generally a $D_{2p}$-brane has a direct
coupling to a $RR$ $F_{(2p+1)}$-brane in IIA theory. Thus for the
case at hand, the magnetic $D6$ and $\bar{D6}$-brane content of
the solution in eq.(21) couple directly to the magnetic $RR$
1-form potential of the $F7$-brane content extracted in eq.(22)
and as a result experience static Coulomb-type force that
eventually keeps the $D6-\bar{D6}$ pair apart against the
gravitational and gauge attractions. \\
Lastly, we see if the conical singularities which were inevitably present in the
$D6-\bar{D6}$ seed solution can now be eliminated by the
introduction of this magnetic $F7$-brane content. To do so, notice
that in this $(D6-\bar{D6})||F7$ case,
$\psi^{\mu}\psi_{\mu}=g_{\tilde{\phi}\tilde{\phi}}=0$ has roots at
the locus of $r=r_{+}$ as well as along the semi-infinite axes
$\theta =0, \pi$. Thus we need to worry about the possible
occurrence of conical singularities both along $\theta=0, \pi$ and
at $r=r_{+}$ again. Assuming that the azimuthal angle coordinate
$\tilde{\phi}$ is identified with period $\Delta \tilde{\phi}$,
the conical deficit along the axes $\theta =0, \pi$ and along the
segment $r=r_{+}$ are given respectively by
\beqa
\delta_{(0,\pi)} &=& 2\pi - \arrowvert {\Delta \tilde{\phi} d\sqrt{g_{\tilde{\phi}\tilde{\phi}}} \over
\sqrt{g_{\theta\theta}}d\theta} \arrowvert_{\theta =0, \pi} = 2\pi - \Delta \tilde{\phi}, \\
\delta_{(r=r_{+})} &=& 2\pi - \arrowvert {\Delta \tilde{\phi} d\sqrt{g_{\tilde{\phi}\tilde{\phi}}} \over
\sqrt{g_{rr}}dr}\arrowvert_{r=r_{+}} = 2\pi - \left[{r_{+}-m \over B(r^2_{+}-a^2)+a}\right]
\Delta \tilde{\phi} \nonumber
\eeqa
where, in this time, we used the $(D6-\bar{D6})||F7$ metric solution given in eq.(21).
Therefore, by choosing $\Delta \tilde{\phi} = 2\pi$ and ``tuning'' the strength of the external magnetic field as
\beq
B = {(r_{+}-m)-a \over (r^2_{+}-a^2)} = {\sqrt{m^2+a^2}-a \over 2mr_{+}}
\eeq
one now can remove all the conical singularities. As stated
earlier, this removal of conical singularities by properly tuning
the strength of the magnetic field amounts to suspending the
$D6-\bar{D6}$ pair in an (unstable) equilibrium configuration by
introducing a force exerted by this magnetic field (i.e., the $RR$
$F7$-brane) to counterbalance the combined gravitational and gauge
attractive force. To see this in a qualitative manner \cite{sen2}, recall first that, when they are well separated,
the distance between $D6$ and $\bar{D6}$ is given roughly by $\sim 2a$ as shown in eq.(7) and in this
large-$a$ limit, the magnetic field strength given above in eq.(24) is $B\simeq m/4a^2$.
Next, since both the gravitational and $RR$ gauge attractive forces between the branes would be
given by $m^2/(2a)^2$ (where we used the fact that the $RR$-charge of a $D6$-brane behaves like
$q\rightarrow m$ for large inter-brane separation as discussed earlier), the total attractive force
goes like $m^2/2a^2$. Thus this combined attractive force would be counterbalanced by the repulsive
force on the magetic dipole of the $D6-\bar{D6}$ pair, $2qB \simeq 2m(m/4a^2) = m^2/2a^2$ provided 
by the properly tuned magnetic field strength $B$ of $RR$ $F7$-brane give above in eq.(24).  

\subsection{$D0-\bar{D0}$ pair supported by $RR$ $F7$-brane}

For the case of $D0-\bar{D0}$ system, which is ``electrically'' $RR$-charged,
it may seem irrelevant to attempt to intersect it with magnetic $F7$-brane to
begin with. As we shall see later on in the appendix, however, the attempt to intersect it with
electric $RR$ fluxbrane via the twisted KK-reduction of $W-\bar{W}$ system in 
$D=11$ supergravity fails. Thus the only remaining option is to intersect it
with magnetic $F7$-brane instead and see what effect this magnetic fluxbrane 
may have on the $D0-\bar{D0}$ system regarding the stabilization particularly
when the brane and the antibrane are brought close to each other. Thus we now 
start with the exact IIA supergravity solution representing the $D0-\bar{D0}$ 
pair which is given, in string frame, by \cite{youm} 
\beqa
ds^2_{10} &=& H^{-1/2}[-dt^2] +
H^{1/2}[\sum^{6}_{m=1} dx^2_{m} + (\Delta+a^2\sin^2 \theta)\left({dr^2\over
\Delta}+d\theta^2\right) + \Delta \sin^2 \theta d\phi^2], \nonumber \\ 
e^{2\phi} &=& H^{3/2}, \\ 
A_{[1]} &=& \left[{2ma\cos \theta \over \Sigma}\right]dt, ~~~F_{[2]}=dA_{[1]} 
\nonumber 
\eeqa 
where again $H(r) = \Sigma /({\Delta + a^2\sin^2 \theta})$. 
Then, as usual, by uplifting this $D0-\bar{D0}$ solution in $D=10$ IIA theory to $D=11$,
we can arrive at the $W-\bar{W}$ (i.e., M-wave/anti-M-wave) solution given by
\beqa
ds^2_{11} &=& e^{-{2\over 3}\phi}ds^2_{10} + e^{{4\over 3}\phi}(dy+A_{\mu}dx^{\mu})^2 \nonumber \\
&=& - H^{-1}dt^2 + H\left(dy + {2ma\cos \theta \over \Sigma}dt\right)^2   \\
&+& [\sum^{6}_{m=1} dx^2_{m} + (\Delta+a^2\sin^2 \theta)\left({dr^2\over
\Delta}+d\theta^2\right) + \Delta \sin^2 \theta d\phi^2]. \nonumber
\eeqa
Now, in order to construct the supergravity solution representing $D0-\bar{D0}$ pair
parallely intersecting with a $RR$ $F7$-brane by introducing the $RR$ $F7$-brane content 
into the $D0-\bar{D0}$ solution given in eq.(25), we, as usual, proceed to perform a 
{\it twisted} KK-reduction on this M-theory $W-\bar{W}$ solution. 
Consider, therefore, performing the non-trivial point identification 
\beq 
(y, ~\phi) \equiv (y + 2\pi n_{1}R, ~\phi + 2\pi n_{1}RB + 2\pi n_{2}) 
\eeq
(with $n_{1}, ~n_{2}\in Z$) on the M-theory $W-\bar{W}$ solution
in eq.(26), followed by the associated skew KK-reduction along the
orbit of the Killing field $l = (\partial/\partial y) + B(\partial/\partial \phi )$ 
where $B$ is again a magnetic field parameter. And this amounts to introducing the
{\it adapted} coordinate $\tilde{\phi} = \phi - By$ which is constant along the orbits 
of $l$ and possesses standard period of $2\pi$ and then proceeding with the standard 
KK-reduction along the orbit of $(\partial/\partial y)$. The result is
\beqa
ds^2_{10} &=& \tilde{\Lambda}^{1/2}\left\{-H^{-1/2}dt^2 + H^{1/2}[\sum^{6}_{m=1}dx^2_{m} +
(\Delta + a^2\sin^2 \theta)\left({dr^2\over \Delta} + d\theta^2\right)]\right\} \nonumber \\
&+& \tilde{\Lambda}^{-1/2}H^{1/2}\Delta \sin^2 \theta\left\{B^2{(2ma\cos\theta)^2\over \Sigma^2}
dt^2 + d\tilde{\phi}^2 - B{4ma\cos\theta \over \Sigma}dtd\tilde{\phi}\right\}, \nonumber \\
e^{{4\over 3}\phi} &=& H\tilde{\Lambda}, \\
A_{[1]} &=& \tilde{\Lambda}^{-1}{2ma\cos\theta \over \Sigma}dt + \tilde{\Lambda}^{-1}H^{-1}
B\Delta \sin^2 \theta d\tilde{\phi}, \nonumber \\
F_{[2]} &=& (\partial_{r}A_{t})dr\wedge dt +
(\partial_{\theta}A_{t})d\theta\wedge dt +
(\partial_{r}A_{\tilde{\phi}})dr\wedge d\tilde{\phi} +
(\partial_{\theta}A_{\tilde{\phi}})d\theta\wedge d\tilde{\phi},
~~~{\rm where} \nonumber \\
\tilde{\Lambda} &=& (1 + H^{-1}B^2\Delta \sin^2 \theta). \nonumber
\eeqa
This solution can be identified with a $D0-\bar{D0}$ pair
parallely intersecting with a magnetic $RR$ $F7$-brane since for
$B=0$, it reduces to the $D0-\bar{D0}$ solution in eq.(25) while
for $m=0$ and $a=0$, it reduces to a $RR$ $F7$-brane solution in
IIA theory given in eq.(22). 
Again, we see if the conical singularities which were present in the
$D0-\bar{D0}$ seed solution can now be eliminated by the
introduction of this magnetic $F7$-brane content. To do so, notice
that in this $(D0-\bar{D0})||F7$ case,
$\psi^{\mu}\psi_{\mu}=g_{\tilde{\phi}\tilde{\phi}}=0$ still has roots at
the locus of $r=r_{+}$ as well as along the semi-infinite axes
$\theta =0, \pi$. Thus we need to worry about the possible
occurrence of conical singularities both along $\theta=0, \pi$ and
at $r=r_{+}$ again. Assuming that the azimuthal angle coordinate
$\tilde{\phi}$ is identified with period $\Delta \tilde{\phi}$,
the conical deficit along the axes $\theta =0, \pi$ and along the
segment $r=r_{+}$ are given respectively by
\beqa
\delta_{(0,\pi)} &=& 2\pi - \arrowvert {\Delta \phi
d\sqrt{g_{\phi\phi}} \over \sqrt{g_{\theta\theta}}d\theta}
\arrowvert_{\theta =0, \pi} = 2\pi - \Delta \phi, \\
\delta_{(r=r_{+})} &=& 2\pi - \arrowvert {\Delta \phi
d\sqrt{g_{\phi\phi}} \over \sqrt{g_{rr}}dr} \arrowvert_{r=r_{+}} =
2\pi - \left(1 + {m^2\over a^2}\right)^{1/2}\Delta \phi \nonumber
\eeqa 
where, of course, we used the $(D0-\bar{D0})||F7$ metric solution
given in eq.(28). To our dismay, but indeed as had been expected due to the
reason stated earlier, the conical singularity structure essentially remains the
same as that for the $D0-\bar{D0}$ system, despite the introduction of the $RR$
$F7$-brane content into the system aiming at counterbalancing the combined 
gravitational and gauge attractions and hence keeping the system against the collapse.
In other words, we still cannot remove the conical singularities along the axes,
$\theta = 0, \pi$ and along the segment $r = r_{+}$ at the same time. Certainly,
this discouraging result demands physical explanation and indeed it can be attributed 
to the fact $D0$ (and $\bar{D0}$) does not couple directly to the flux of $RR$ $F7$-brane
and as a result experiences no Coulomb-type force from its presence.
As we mentioned earlier, a $D_{p}$-brane couples only to the flux of a $F_{(p+1)}$-brane 
and this fact comes from the defining nature of the $RR$ $F_{(p+1)}$-brane \cite{flux} according to
which a $F_{(p+1)}$-brane is a $(p+2)$-dimensional object in the $(8-p)$-dimensional
transverse space. And the core of this $F_{(p+1)}$-brane carries a $(8-p)$-form {\it magnetic}
$RR$ field strength with flux piercing the transverse space. Therefore, only the 
$D6-\bar{D6}$ system, which has non-trivial coupling to the $F7$-brane, can be
balanced in an (unstable) equilibrium against the combined gravitational and gauge attractions.
The others, such as $D0-\bar{D0}$ we just discussed and $D2-\bar{D2}$ and $D4-\bar{D4}$, the 
case of which can be examined in prcisely the same manner, cannot be stabilized via the
introduction of the magnetic $F7$-brane simply because they do not have non-trivial coupling.
For the cases of $D0-\bar{D0}$ and $D2-\bar{D2}$ systems, however, it may seem irrelevant to
intersect them with magnetic $F7$-brane to begin with since these configuration are 
{\it electrically} $RR$ charged. But as will be demonstrated later on in the appendix, 
any attempt to intersect 
them with {\it electric} $RR$ fluxbrane via the twisted KK-reduction from M-theory $W-\bar{W}$
and $M2-\bar{M2}$ systems respectively in D=11 fails. Thus we are forced to intersect them 
with magnetic $F7$-branes instead and see what effect this magnetic $F7$-brane may have on 
$D0-\bar{D0}$ or $D2-\bar{D2}$ system concerning the stabilization particularly when they 
are brought closer and closer to each other. It turns out that there indeed is a non-trivial
effect which is puzzling at first sight but admits convincing interpretation on second thought.
To discuss it in great detail, we first remind our earlier observation that the parameter ``$a$''
appearing in the supergravity solutions representing  $D_{2p}-\bar{D}_{2p}$ systems in IIA theory
can be thought of as representing the proper separation between the brane and the antibrane all
the way to the zero distance.
We now take the $D0-\bar{D0}$ case which is under consideration and see what happens as 
the brane and the antibrane approach each other, namely as $a\rightarrow 0$. First, in the
absence of the magnetic $RR$ $F7$-brane content, the $D0-\bar{D0}$ solution in the limit 
$a\rightarrow 0$ becomes
\beqa
ds^2_{10} &=& \left(1 - {2m\over r}\right)^{1/2}[-dt^2] +
\left(1 - {2m\over r}\right)^{-1/2}[\sum^{6}_{m=1} dx^2_{m} + dr^2
+ r^2\left(1 - {2m\over r}\right)(d\theta^2 + \sin^2 \theta d\phi^2)], \nonumber \\ 
e^{2\phi} &=& \left(1 - {2m\over r}\right)^{-3/2}, ~~~A_{[1]} = 0  
\eeqa 
where we used $\Sigma \rightarrow r^2$, $\Delta \rightarrow r^2(1-2m/r)$, and hence 
$H \rightarrow (1-2m/r)^{-1}$ as $a\rightarrow 0$. In this limit, the opposite $RR$ charges
carried by $D0$ and $\bar{D0}$ annihilated each other since $A_{[1]}=0$ and the solution
now has the topology of $R\times R^{7}\times S^{2}$. Particularly, the $SO(3)$-isometry in
the transverse space implies that, as they approach, $D0$ and $\bar{D0}$ actually merge
and as a result a curvature singularity develops at the center $r=0$. On the other hand, 
looking at the solution representing the $D0-\bar{D0}$ pair embedded in the magnetic 
$F7$-brane, it does not appear to be possible to bring $D0$ and $\bar{D0}$ close enough
to make them merge completely. Thus to see this, consider the $a\rightarrow 0$ limit of
the $(D0-\bar{D0})||F7$ solution
\beqa
ds^2_{10} &=& \tilde{\Lambda}^{1/2}\left\{\left(1 - {2m\over r}\right)^{1/2}[-dt^2] +
\left(1 - {2m\over r}\right)^{-1/2}[\sum^{6}_{m=1} dx^2_{m} + dr^2
+ r^2\left(1 - {2m\over r}\right)d\theta^2]\right\} \nonumber \\
&+& \tilde{\Lambda}^{-1/2}r^2\left(1 - {2m\over r}\right)^{1/2}\sin^2 \theta 
d\tilde{\phi}^2,  \\ 
e^{2\phi} &=& \tilde{\Lambda}^{3/2}\left(1 - {2m\over r}\right)^{-3/2}, ~~~A_{[1]} = 
\tilde{\Lambda}^{-1}\left(1 - {2m\over r}\right)^{2}Br^2\sin^2 \theta d\tilde{\phi}
\nonumber  
\eeqa 
where now $\tilde{\Lambda} = [1 + B^2r^2(1-2m/r)^{2}\sin^2\theta]$ and $A_{[1]}$ is the
magnetic vector potential for the $F7$-brane. It is now obvious that for finite $B\neq 0$,
the portion of the transverse space fails to exhibit $SO(3)$-isometry. Instead, the 
solution now possesses {\it axisymmetry}, namely, the metric solution has explicit $\theta$-dependence
coming from the factor $\tilde{\Lambda}$ and this is the manifestation that even for very
small separation, the two brane configuration structure still persists. In fact, the
axisymmetry itself even in the limit $a\rightarrow 0$ of the solution is no surprise as it has
been expected to some extent since the $(D0-\bar{D0})||F7$ solution involves the axisymmetric
$F7$-brane content from the outset. Rather the point is that, in the $a\rightarrow 0$ limit of
the $(D0-\bar{D0})||F7$ solution given above, one never knows whether this axisymmetry
comes from the remaining $F7$-brane content after the complete merging of the 
$D0-\bar{D0}$ pair or from the surviving brane-antibrane configuration so long as one keeps
the non-vanishing content of the branes, i.e., $m\neq 0$. Thus generically, one should 
regard that even in the limit $a\rightarrow 0$, the two brane structure may have a good
chance to survive. Of course, if we
turn off the $F7$-brane content, i.e., if we set $B = 0$, then for $a\rightarrow 0$, 
$D0$ and $\bar{D0}$ merge completely as they should. This observation indicates that 
although the magnetic $F7$-brane and $D0$ (and $\bar{D0}$) do not directly couple and
hence $F7$ fails to serve to stabilize the $D0-\bar{D0}$ system against the eventual 
collapse for finite separation, when $D0$ and $\bar{D0}$ are brought close enough together,
the $F7$-brane turns out to play the role of keeping them  from annihilating each other
completely. And it is rather straightforward to see that the same is true for $D2-\bar{D2}$
and $D4-\bar{D4}$ systems as well (whose supergravity solutions are known \cite{youm}). 
Within the context of the supergravity analysis, this
picture is an apparent puzzle and demands some resolution. As was mentioned earlier in
the introduction, one may naturally expect that the simplest endpoints of the semi-classical
instability of the $D_{2p}-\bar{D}_{2p}$ systems would be a supersymmetric vacuum. And 
since the introduction of $RR$ $F7$-brane content cannot remove the semi-classical instability
of the $D_{2p}-\bar{D}_{2p}$ systems, one may still expect that they should eventually
merge completely when they approach each other even in the presence of the fluxbrane.
But rather to our surprise, this turned out not to be the case. Indeed, the possible answer
to this puzzle may lie in the validity of the semi-classical supergravity description of the
system. Namely, the supergravity solutions representing $(D_{2p}-\bar{D}_{2p})||F7$ systems
cannot be trusted for stability analysis all the way down to $a\rightarrow 0$ and obviously
they invalidate as the separation between the branes approaches the string length scale, i.e.,
$a \leq \sqrt{\alpha'} = l_{s}$. Put differently, for very small separation of order
$a \sim \sqrt{\alpha'} = l_{s}$, the supergravity description of the system breaks down and
we should in principle employ stringy analysis instead in terms of tachyonic mode arising for 
$a \leq l_{s}$ in the spectrum of open strings stretched between $D_{p}$ and $\bar{D}_{p}$. 
Firstly, as for the
quantum instability associted with the non-BPS $D_{p}-\bar{D}_{p}$ system, it is by now widely
accepted that for a nearly coincident  $D_{p}-\bar{D}_{p}$ pair, the spectrum of open strings
connecting $D_{p}$ and $\bar{D}_{p}$ develops a tachyonic mode and this open string tachyon 
``condenses'' as $D_{p}$ and $\bar{D}_{p}$ annihilate each other to become a supersymmetric 
vacuum or evolve to produce a stable lower-dimensional brane. And it has been estimated by Callan
and Maldacena \cite{malda} that the typical time scale for this $D_{p}-\bar{D}_{p}$ annihilation process is
of order $1/\sqrt{g_{s}}$ where $g_{s}$ denotes the fundamental string couplng,
$g_{s} = e^{\phi(\infty)}$. Secondly, as for the quantum instability associated with a $RR$
fluxbrane, it has been conjectured and generally believed that it would presumably be linked to
the closed string tachyonic mode. We now elaborate on this last point. As is well-known, one of
the simplest ways to construct a $RR$ $F7$-brane is via the ``twisted'' KK-reduction of the
$D=11$ Minkowski spacetime, which is a M-theory vacuum. As such, the $F7$-brane breaks all the
supersymmetries and hence should be unstable and decay. 
Indeed, it has been known for some time \cite{costa, flux}
that the Melvin-type fluxtube universe like the $F7$-brane actually decays at the rate given by
$\Gamma \sim e^{-I}$, with ``$I$'' being the Euclidean instanton action and the instanton
configuration related to this decay of the Melvin-type magnetic fluxtube universe is the
Euclidean Kerr geometry in an arbitrary dimension. And it is generally expected that the endpoint 
of this $RR$ $F7$-brane decay would be either a supersymmetric closed string vacuum or the
nucleation of the $D6-\bar{D6}$ pair via the brainy Schwinger process \cite{costa}. 
Particularly, it has been
conjectured that the fluxbrane decay to a supersymmetric vacuum should be linked to the {\it closed}
string tachyon condensation since it involves the decay of the spacetime itself \cite{flux, russo1}. 
For the case at hand,
we have both $D_{p}-\bar{D}_{p}$ pair and the $RR$ $F7$-brane in the system and each is unstable
for the reasons just stated. What is more, it is in many respects evident that the presence of the
magnetic $F7$-brane, i.e., the external magnetic field changes the status of the quantum instabilty
of the $D_{p}-\bar{D}_{p}$ system. Namely, due to the additional energy density introduced by the
external magnetic field (i.e., the $F7$-brane), the total energy density of the 
$(D_{2p}-\bar{D}_{2p})||F7$ system now would be given by
\beq
E_{tot} = V(T) + 2M_{D} + \epsilon_{F7}
\eeq
where again $V(T)$ and $M_{D}$ are the tachyon potential and the $D$-brane tension respectively and
$\epsilon_{F7}$ denotes the contribution to the total energy density coming from $F7$-brane, i.e., the
magnetic field energy density. Note that the tachyon potential $V(T)$ here would remain unchanged 
from that in the absence of the magnetic $RR$ $F7$-brane as the $NS$-charged open strings stretched
between $D_{2p}$ and $\bar{D}_{2p}$ have no direct coupling to the flux of $RR$ $F7$-brane. Now in
this stringy description, as $D_{2p}$ and $\bar{D}_{2p}$ approach each other, the open string tachyon
field $T$, having essentially the same potential as the one without $F7$, may still condense, i.e., 
its mass squared may evolve from being negative around the false vacuum expectation value (vev) to 
being positive at the true vev, say, $T_{0}$ as it rolls down toward the (negative) minimum of the
potential, $V(T_{0})$. Even when the tachyon reaches the minimum $V(T_{0})$ of its potential, however,
the brane and the antibrane would not necessarily annihilate since the endpoint of this tachyon
condensation is no longer a supersymmetric vacuum but instead it is a left-over $F7$-brane for which
the supersymmetry is completely broken, as can be deduced from the consideration,
$E_{tot} = V(T_{0}) + 2M_{D} + \epsilon_{F7} = \epsilon_{F7}$ (where we used Sen's conjecture \cite{sen1} 
in the absence of the $F7$-brane content, $V(T_{0}) + 2M_{D} = 0$).
Of course, there is another possibility in which the (negative) $V(T_{0})$ cancels instead with
the part or all of $\epsilon_{F7}$ rather than it does exactly with $2M_{D}$. In this alternative situation, upon
the tachyon condensation the left-over would be something that again breaks the supersymmetry completely.
In other words, the brane-antibrane system would not be driven to the complete spontaneous
annihilation as the endpoint of the tachyon condensation does not, in any case,
enhance any supersymmetry of the 
system. As a result, as long as the $F7$-brane is there, $D_{p}$ and $\bar{D}_{p}$ 
would not necessarily annihilate each other via the open string tachyon condensation
and this quantum perspective is indeed consistent with the result of semi-classical supergravity
analysis given earlier in which it has been demonstrated that $D_{2p}-\bar{D}_{2p}$ pairs may not
necessarily 
merge and annihilate in the presence of the $F7$-brane content even if they are brought close
enough together. Besides, this argument to resolve the puzzling role played by the $RR$ 
$F7$-brane content in the $D_{2p}-\bar{D}_{2p}$ pairs $(p=2, 1, 0)$ holds true for the
case of $D6-\bar{D6}$ system as well although there, the $RR$ $F7$-brane has actually
the direct coupling to the $D6-\bar{D6}$ system and hence is able to provide the system
with even a classical stability (i.e., an unstable equilibrium) generally for some finite 
separation between the branes. Namely, when $D6$ and $\bar{D6}$ are brought close enough to
each other, the branes can be supported only until the $F7$-brane itself disappears by 
decaying to a vacuum or to $D6-\bar{D6}$ pairs via brany Schwinger process \cite{costa}.
And this indicates that after all, the result of supergravity analysis given earlier  
was not totally wrong although it should not be naively trusted. $F7$-brane, however, is itself
unstable (as it breaks all the supersymmetries) and hence decays eventually. Therefore, the overall
picture of the quantum instability of the $(D_{2p}-\bar{D}_{2p})||F7$ systems can be stated as
follows : \\
{\it Within the time scale for the decay of magnetic $F7$-brane, $D_{p}-\bar{D}_{p}$ pair would be
supported against collapse and the subsequent annihilation. Once $F7$-brane itself decays, 
$D_{p}$ and $\bar{D}_{p}$ would now annihilate each other presumably leaving supersymmetric vacuum 
behind as the tachyonic mode in the spectrum of the open strings stretched between 
$D_{p}$ and $\bar{D}_{p}$ condenses. Namely, the presence of the $RR$ $F7$-brane just
``delays'' the annihilation process of $D_{p}-\bar{D}_{p}$ pair but can never eliminate the
instability of the $D_{p}-\bar{D}_{p}$ pair completely !} \\
Next, since it is the presence of the $F7$-brane which ``delays'' the annihilation of the
$D_{p}-\bar{D}_{p}$ system, this decay mechanism might deserve closer examination. And it would be
of particular interest to study what the effect of the presence of the $D_{p}-\bar{D}_{p}$ pair   
on the decay rate of the $F7$-brane really is. In order to estimate the $F7$-brane decay rate, all
that is required is to find the instanton mediating the decay with the same
asymptotics as those of the $F7$-brane since the two have to be matched in the asymptotic region.
In the presence of the $F7$-brane alone and nothing else, it was rather straightforward to find the 
associated instanton configuration and that was, as mentioned, the higher-dimensional generalization 
of the Euclidean Kerr metric. And thus the evaluation of the corresponding Euclidean instanton
action, $I(instanton)$ was rather unambiguous, as well. For the case at hand, however, when both
$F7$-brane and $D_{p}-\bar{D}_{p}$ pair are present, things get much more involved. Namely, since
the presence of the $D_{p}-\bar{D}_{p}$ pair changes the asymptotics of the $F7$-brane geometry
in a highly non-trivial fashion as we actually have seen earlier in eq.(21) or in eq.(28), 
it would be practically
almost impossible to find the associated instanton configuration having this complicated
asymptotics. And this, in turn, indicates that now the explicit evaluation of the corresponding 
Euclidean instanton action would not be available, either. Thus, one can only hope to determine
whether the instanton action (decay rate) gets smaller (higher) or else it is the other way around.
And the clue that would lead us to the right answer to this question is undoubtedly linked to 
whether or not the presence of the $D_{p}-\bar{D}_{p}$ pair would add more instability to the
$F7$-brane accelerating its decay process. Unfortunately, any conclusive statement concerning this
point appears to be beyond our reach for the moment, but our best guess is that presumably,
the presence of the $D_{p}-\bar{D}_{p}$ pair would increase the instability of the $F7$-brane
and hence elevate its decay rate. And this guess is based upon the fact that the 
$D_{p}-\bar{D}_{p}$ system is itself an unstable non-BPS configuration involving the open string
tachyonic mode. 

\section{$NS-anti-NS$ systems supported by $NSNS$ fluxbrane ?}\label{ }

Thus far we have considered the intersection of non-BPS $D_{2p}-\bar{D}_{2p}$ systems with the
magnetic $RR$ $F7$-brane in IIA theory in order to study the role played by the $RR$ $F7$-brane
concerning the semi-classical and quantum (in terms of open string tachyon condensation)
instability of the $D_{2p}-\bar{D}_{2p}$ systems. Now the remaining unstable non-perturbative
spectrum (or non-BPS solutions) of D=10, type IIA supergravity theory are $F1-\bar{F1}$ and
$NS5-\bar{NS5}$ systems. Since these are charged under the $NSNS$ two-form tensor field
$B_{[2]}$ (electrically for $F1-\bar{F1}$ and magnetically for $NS5-\bar{NS5}$), now it would
be natural to attempt to intersect them with the $NSNS$ $F6$-brane and see if the $NSNS$
$F6$-brane can play an analogous role regarding the semi-classical and quantum instability
which is supposed to reside in these systems. At this point, it seems noteworthy that a
(particularly special case of) $NSNS$ $F6$-brane is related to the $RR$ $F7$-brane via the
chain of duality transformations, $U = TST$ with the $T$-duals acting on the same isometry
direction \cite{russo1}. As a result, it is tempting to expect that, for instance, the
$(NS5-\bar{NS5})||(NSNS ~F6)$ solution might as well be related to the 
$(D6-\bar{D6})||(RR ~F7)$ solution via the same $U$-duality transformation just mentioned.
As we shall see in a moment, however, this turns out not to be the case. Just as the
$D6-\bar{D6}$ system alone possesses direct coupling to the $RR$ $F7$-brane but no other
$D_{2p}-\bar{D}_{2p}$, only the $NS5-\bar{NS5}$ but not $F1-\bar{F1}$ has direct coupling
to the $NSNS$ $F6$-brane. Thus one may naturally expect that the $NS5-\bar{NS5}$ system
would exclusively be counterbalanced against the combined gravitational and gauge attractions
by the introduction of the $NSNS$ $F6$-brane content into the system. As the fact that 
$(NS5-\bar{NS5})||(NSNS ~F6)$ solution is not really related via the $U$-duality to the
$(D6-\bar{D6})||(RR ~F7)$ solution already signals, this naive expectation turns out not to
hold either. Thus in the following, we shall discuss this rather puzzling issue in some detail
and attempt to provide a relevant resolution.    

\subsection{$NS5-\bar{NS5}$ pair intersecting with $NSNS$ $F6$-brane}

We now start with the exact IIA supergravity solution representing the $NS5-\bar{NS5}$ 
pair which is given, in string frame, by \cite{youm} 
\beqa
ds^2_{10} &=& -dt^2 + \sum^{5}_{i=1}dx^2_{i} +
H[dx^2_{6} + (\Delta+a^2\sin^2 \theta)\left({dr^2\over
\Delta}+d\theta^2\right) + \Delta \sin^2 \theta d\phi^2], \nonumber \\ 
e^{2\phi} &=& H, \\ 
B_{[2]} &=& \left[{2mra\sin^2 \theta \over \Delta + a^2 \sin^2 \theta}\right]dx_{6}\wedge d\phi, 
\nonumber \\
H_{[3]} &=& dB_{[2]} \nonumber \\ 
&=& {{2mra\sin^2 \theta (r^2 + a^2\cos^2 \theta)}\over {(\Delta + a^2 \sin^2 \theta)^2}}
dx_{6}\wedge dr\wedge d\phi -
{{4mra\sin \theta \cos \theta \Delta}\over {(\Delta + a^2 \sin^2 \theta)^2}}
dx_{6}\wedge d\theta \wedge d\phi  \nonumber 
\eeqa 
where again, $H(r) = \Sigma /({\Delta + a^2\sin^2 \theta})$. 
Then, as usual, by using the M/IIA duality, we uplift this $NS5-\bar{NS5}$ solution in $D=10$ 
IIA theory to $D=11$, to get the $M5-\bar{M5}$ solution given by
\beqa
ds^2_{11} &=& e^{-{2\over 3}\phi}ds^2_{10} + e^{{4\over 3}\phi}(dy+A_{\mu}dx^{\mu})^2 
\nonumber \\
&=& H^{-1/3}[-dt^2 + \sum^{5}_{i=1}dx^2_{i}] + H^{2/3}[dx^2_{6} + dx^2_{7} + 
(\Delta+a^2\sin^2 \theta)\left({dr^2\over \Delta}+d\theta^2\right) + \Delta \sin^2 \theta d\phi^2], 
\nonumber \\
F^{11}_{[4]} &=& {{2mra\sin^2 \theta (r^2 + a^2\cos^2 \theta)}\over {(\Delta + a^2 \sin^2 \theta)^2}}
dx_{6}\wedge dx_{7}\wedge dr\wedge d\phi \\
&-& {{4mra\sin \theta \cos \theta \Delta}\over {(\Delta + a^2 \sin^2 \theta)^2}}
dx_{6}\wedge dx_{7}\wedge d\theta \wedge d\phi.  \nonumber 
\eeqa
We first identify the coordinate on the M-theory circle as $y=x_{6}$. Then, consider taking the
quotient of this $(M5-\bar{M5})$ spacetime, namely identifying points along the orbit of the
Killing field $l = (\partial/\partial y) + B(\partial/\partial \phi)$, i.e.,
\beq
(y, ~\phi)\equiv (y + 2\pi n_{1}R, ~\phi + 2\pi n_{1}RB + 2\pi n_{2}) 
\eeq
(with $n_{1}, ~n_{2}\in Z$). This amounts to introducing the ``adapted'' coordinate
$\tilde{\phi} = \phi - By$ which is constant along the orbits of $l$ and possesses standard period
of $2\pi$, i.e.,
\beqa
ds^2_{11} &=& H^{-1/3}[-dt^2 + \sum^{5}_{i=1}dx^2_{i}] \\
&+& H^{2/3}[dx^2_{7} + (\Delta+a^2\sin^2 \theta)\left({dr^2\over \Delta}+d\theta^2\right) 
+ \Delta \sin^2 \theta (d\tilde{\phi} + Bdy)^2 + dy^2]. \nonumber
\eeqa  
Earlier, when constructing the $RR$ $F7$-brane, we performed the usual KK-reduction along
$(\partial/\partial y)$. But in this time, consider performing the the KK-compactification
along the orbit of $(\partial/\partial x_{7})$ instead, i.e.,
\beq
ds^2_{11} = e^{-{2\over 3}\phi}ds^2_{10} + e^{{4\over 3}\phi}(dx_{7}+A_{\mu}dx^{\mu})^2
\eeq
to get the 10-dimensional fields
\beqa
ds^2_{10} &=& -dt^2 + \sum^{5}_{i=1}dx^2_{i} +
H[(\Delta+a^2\sin^2 \theta)\left({dr^2\over \Delta}+d\theta^2\right) + 
\Delta \sin^2 \theta (d\tilde{\phi} + Bdy)^2 + dy^2], \nonumber \\ 
e^{{4\over 3}\phi} &=& H^{2/3}, ~~~A_{[1]} = 0.
\eeqa
In addition, from  
\beq
F^{11}_{[4]} = F^{IIA}_{[4]} + H_{[3]}\wedge dx_{7},
\eeq
we also get
\beqa
F^{IIA}_{[4]} &=& 0, \\
H_{[3]} &=& {{2mra\sin^2 \theta (r^2 + a^2\cos^2 \theta)}\over {(\Delta + a^2 \sin^2 \theta)^2}}
dy\wedge dr\wedge d\tilde{\phi} -
{{4mra\sin \theta \cos \theta \Delta}\over {(\Delta + a^2 \sin^2 \theta)^2}}
dy\wedge d\theta \wedge d\tilde{\phi}  \nonumber 
\eeqa
which is the 3-form magnetic $NSNS$ field strength sourced by the $NS5-\bar{NS5}$ pair, since
the associated 2-form magnetic $NSNS$ tensor potential is given by
\beq
B_{[2]} = \left[{2mra\sin^2 \theta \over \Delta + a^2 \sin^2 \theta}\right]dy\wedge d\tilde{\phi}.
\eeq
Thus this new solution can be identified with a $NS5-\bar{NS5}$ system intersecting with a 
$NSNS$ $F6$-brane since for $B=0$, it reduces to the usual $NS5-\bar{NS5}$ solution given
earlier, while for $m=0$ and $a=0$, it reduces to (a special case of) $NSNS$ $F6$-brane.
To see this last point, set $m=0$ and $a=0$ in the solution given above to get
\beqa
ds^2_{10} &=& -dt^2 + \sum^{5}_{i=1}dx^2_{i} + dr^2 + r^2 d\theta^2 + r^2\sin^2 \theta
(d\tilde{\phi} + Bdx_{9})^2 + dx^2_{9}, \nonumber \\
e^{2\phi} &=& 1, ~~~B_{[2]} = 0
\eeqa
where we set $y=x_{9}$. Note first that this is indeed a special case of the more
general $NSNS$ $F6$-brane solution \cite{russo1} of $D=10$ type IIA theory. Certainly, this solution
is locally-flat but it may have non-trivial topology. To see this, recall that this metric
solution can be formally obtained from the flat metric via the shift $\phi \rightarrow
\phi + Bx_{9}$. As a result, a shift of $x_{9}$ by the period of the compactification 
circle $2\pi R$ induces rotaion in the transverse plane by $2\pi RB$. Thus this metric becomes
topologically non-trivial if $BR \neq n$ $(n\in Z)$. It is interesting to note \cite{russo1}, however,
that even for the topologically trivial case, if, particularly, $BR = 2k + 1$ $(k=0, \pm 1,...)$,
the {\it superstring} theory on this background is still non-trivial (i.e., not equivalent to 
that on the flat spacetime) since the spacetime fermions change its sign under $2\pi$ rotation
in the transverse plane. And this means that the $BR = 1$ case (indeed all cases with 
$BR = 2k+1$ are equivalent) represents a superstring with {\it antiperiodic} fermionic boundary
condition in $x_{9}$-direction.  And the magnetic $NSNS$ fluxbrane
content in the new solution above becomes noticeable if one carries out one more time of
dimensional reduction of this solution in eq.(38) along $(\partial/\partial x_{9})$, i.e.,
\beqa
ds^2_{10} &=& -dt^2 + \sum^{5}_{i=1}dx^2_{i} +
H[(\Delta+a^2\sin^2 \theta)\left({dr^2\over \Delta}+d\theta^2\right) + 
\Delta \sin^2 \theta (d\tilde{\phi} + Bdx_{9})^2 + dx^2_{9}] \nonumber \\   
&=& ds^2_{9} + e^{2\phi}(dx_{9}+A_{\alpha}dx^{\alpha})^2.
\eeqa
Then the resulting 9-dimensional fields are
\beqa
ds^2_{9} &=& -dt^2 + \sum^{5}_{i=1}dx^2_{i} +
H[(\Delta+a^2\sin^2 \theta)\left({dr^2\over
\Delta}+d\theta^2\right) + f^{-1}\Delta \sin^2 \theta d\tilde{\phi}^2], \nonumber \\ 
e^{2\phi} &=& Hf, \\ 
A_{[1]} &=& f^{-1}B\Delta \sin^2 \theta d\tilde{\phi} \nonumber
\eeqa
with $f=(1+B^2 \Delta \sin^2 \theta)$. Evidently, the emergence of the KK gauge field 
$A_{\tilde{\phi}}d\tilde{\phi}$ indicates that this 9-dimensional supergravity solution
and hence its 10-dimensional ancestor given earlier indeed represent Melvin-type magnetic
fluxbrane.

\subsection{$F1-\bar{F1}$ pair intersecting with $NSNS$ $F6$-brane}

Since $F1-\bar{F1}$ system is ``electrically'' $NSNS$ charged, it would seem natural to attempt
to intersect it with an electric $NSNS$ fluxbrane. Such an attempt, via the twisted KK-reduction 
of $M2-\bar{M2}$ system, however, fails again since the electric fluxbrane constructed in this
manner turns out to be trivial having a null structure just as what happens when one attempts
to intersect a $D0-\bar{D0}$ system with an electric $RR$ fluxbrane that we discussed earlier.
Thus we attempt to intersect the $F1-\bar{F1}$ system with the magnetic $NSNS$ $F6$-brane instead
and examine what effect this magnetic fluxbrane may have on the classical and quantum instability
of the $F1-\bar{F1}$ system. \\
We now start with the exact IIA supergravity solution representing the $F1-\bar{F1}$ 
pair which is given, in string frame, by \cite{youm} 
\beqa
ds^2_{10} &=& H^{-1}[-dt^2 + dx^2_{1}] + \sum^{6}_{m=2}dx^2_{m} +
(\Delta+a^2\sin^2 \theta)\left({dr^2\over
\Delta}+d\theta^2\right) + \Delta \sin^2 \theta d\phi^2, \nonumber \\ 
e^{2\phi} &=& H^{-1}, \\ 
B_{[2]} &=& -\left[{2ma\cos \theta \over \Sigma}\right]dt\wedge dx_{1}
\nonumber 
\eeqa 
where again, $H(r) = \Sigma /({\Delta + a^2\sin^2 \theta})$. 
Then by using the M/IIA duality, we uplift this $F1-\bar{F1}$ solution in $D=10$ 
IIA theory to $D=11$, to get the $M2-\bar{M2}$ solution given by
\beqa
ds^2_{11} &=& H^{-2/3}[-dt^2 + \sum^{2}_{i=1}dx^2_{i}] + H^{1/3}[\sum^{7}_{m=3}dx^2_{m} + 
(\Delta+a^2\sin^2 \theta)\left({dr^2\over \Delta}+d\theta^2\right) + \Delta \sin^2 \theta d\phi^2], 
\nonumber \\
A^{11}_{[3]} &=& -\left[{2ma\cos \theta \over \Sigma}\right]dt\wedge dx_{1}\wedge dx_{2}. 
\eeqa
Now, we first identify the coordinate on the M-theory circle as $y=x_{7}$ and consider taking the
quotient of this $(M2-\bar{M2})$ spacetime, namely identifying points along the orbit of the
Killing field $l = (\partial/\partial y) + B(\partial/\partial \phi)$, i.e.,
\beq
(y, ~\phi)\equiv (y + 2\pi n_{1}R, ~\phi + 2\pi n_{1}RB + 2\pi n_{2}) 
\eeq
which amounts to introducing the ``adapted'' coordinate
$\tilde{\phi} = \phi - By$ in terms of which the metric for the $M2-\bar{M2}$ solution is 
rewritten as
\beqa
ds^2_{11} &=& H^{-2/3}[-dt^2 + \sum^{2}_{i=1}dx^2_{i}] \\
&+& H^{1/3}[\sum^{6}_{m=3}dx^2_{m}+ (\Delta+a^2\sin^2 \theta)\left({dr^2\over \Delta}+d\theta^2\right) 
+ \Delta \sin^2 \theta (d\tilde{\phi} + Bdy)^2 + dy^2]. \nonumber
\eeqa  
Consider now performing the the KK-compactification
along the orbit of $(\partial/\partial x_{2})$, i.e.,
\beq
ds^2_{11} = e^{-{2\over 3}\phi}ds^2_{10} + e^{{4\over 3}\phi}(dx_{2}+A_{\mu}dx^{\mu})^2
\eeq
to get the 10-dimensional fields
\beqa
ds^2_{10} &=& H^{-1}[-dt^2 + dx^2_{1}] + \sum^{6}_{m=3}dx^2_{m} +
(\Delta+a^2\sin^2 \theta)\left({dr^2\over \Delta}+d\theta^2\right) \nonumber \\
&+& \Delta \sin^2 \theta (d\tilde{\phi} + Bdy)^2 + dy^2, \\
e^{{4\over 3}\phi} &=& H^{-2/3}, ~~~A_{[1]} = 0. \nonumber
\eeqa
In addition, from  
\beq
A^{11}_{[3]} = A^{IIA}_{[3]} + H_{[2]}\wedge dx_{2},
\eeq
we also get
\beqa
A^{IIA}_{[3]} = 0, 
~~~B_{[2]} = -\left[2ma\cos \theta \over \Sigma\right]dt\wedge dx_{1} 
\eeqa
which precisely is the 2-form electric $NSNS$ tensor potential sourced by a $F1-\bar{F1}$ pair.
Thus this new solution can be identified with a $F1-\bar{F1}$ system intersecting with a 
$NSNS$ $F6$-brane since for $B=0$, it correctly reduces to the usual $F1-\bar{F1}$ solution given
earlier, while for $m=0$ and $a=0$, it reduces to (a special case of) $NSNS$ $F6$-brane given in
eq.(42). And next, the magnetic $NSNS$ fluxbrane
content in this new solution above becomes recognizable by reducing one more time 
down to 9-dimensions along $(\partial/\partial x_{9})$, i.e.,
\beqa
ds^2_{10} &=& H^{-1}[-dt^2 + dx^2_{1}] + \sum^{5}_{m=2}dx^2_{m} +
(\Delta+a^2\sin^2 \theta)\left({dr^2\over \Delta}+d\theta^2\right) \nonumber \\
&+& \Delta \sin^2 \theta (d\tilde{\phi} + Bdx_{9})^2 + dx^2_{9} 
= ds^2_{9} + e^{2\phi}(dx_{9}+A_{\alpha}dx^{\alpha})^2.
\eeqa
The resulting 9-dimensional fields are
\beqa
ds^2_{9} &=& H^{-1}[-dt^2 + dx^2_{1}] + \sum^{5}_{m=2}dx^2_{m} +
(\Delta+a^2\sin^2 \theta)\left({dr^2\over
\Delta}+d\theta^2\right) + f^{-1}\Delta \sin^2 \theta d\tilde{\phi}^2, \nonumber \\ 
e^{2\phi} &=& f, \\ 
A_{[1]} &=& f^{-1}B\Delta \sin^2 \theta d\tilde{\phi} \nonumber
\eeqa
with $f=(1+B^2 \Delta \sin^2 \theta)$. Again, the emergence of the KK gauge field 
$A_{\tilde{\phi}}d\tilde{\phi}$ indicates that this 9-dimensional supergravity solution
and hence its 10-dimensional ancestor given above do represent Melvin-type magnetic
fluxbrane. \\
It is rather obvious that for both $NS5-\bar{NS5}$ and $F1-\bar{F1}$ systems, the conical
singularity structure remains the same although we have introduced into the system the
$NSNS$ $F6$-brane content again aiming at counterbalancing the gravitational and gauge
attractions and hence keeping the brane-antibrane systems against collision and
subsequent annihilation. Namely, we still cannot eliminate the conical singularities along
the axes $\theta=0, \pi$ and along the segment $r=r_{+}$ at the same time. Moreover, there
appears to be a rather unexpected point that needs to be clarified with care. Earlier, we
discussed the intersection of the $RR$ $F7$-brane with $D_{2p}-\bar{D}_{2p}$ 
pairs in IIA-theory. There, we noticed that since generally a $D_{p}$ brane couples directly
to the flux of a $F_{(p+1)}$-brane, only the $D6-\bar{D6}$ system, but not others, which has
non-trivial coupling to the $RR$ $F7$-brane, can be balanced in an unstable equilibrium against
the combined gravitational ($NSNS$) and gauge ($RR$) attractions. Along this line of argument,
for the case at hand, we may naturally expect that it would be the $NS5-\bar{NS5}$ system,
but obviously not $F1-\bar{F1}$, which directly couples to the magnetic flux of the $NSNS$
$F6$-brane and, as a result, can be balanced in an unstable equilibrium. This, however, turns
out not to be the case. Namely, despite the introduction of the $NSNS$ $F6$-brane content, the
$NS5-\bar{NS5}$ system, let alone the $F1-\bar{F1}$ system, still preserve essentially the same
conical singularity structure and hence exhibit the unaffected semi-classical instability.
Indeed this puzzle has an immediate explanation and it is due to the fact that unlike the
$RR$ $F7$-brane content which carries the non-vanishing magnetic 2-form flux $F_{[2]}$, the $NSNS$
$F6$-brane content as has been constructed via the KK-reduction not along $(\partial/\partial y)$
but instead along $(\partial/\partial x_{9})$ carries no non-trivial magnetic 3-form flux
$H_{[3]}$ proportional to the magnetic strength parameter $B$. We already have witnessed this
point in the expression for the pure $NSNS$ $F6$-brane content that has been extracted by
setting $m=0$ and $a=0$ in the $(NS5-\bar{NS5})||F6$ and $(F1-\bar{F1})||F6$ solutions
given above. As a result, there is simply no magnetic $NSNS$ flux for the $NS5-\bar{NS5}$
pair to couple to and hence no repulsive force between the brane and the antibrane to 
counterbalance the combined gravitational and gauge attractions. Unlike the 
$(D6-\bar{D6})||F7$ and $(D0-\bar{D0})||F7$ systems we discussed earlier, the point worthy
of note in the present $(NS5-\bar{NS5})||F6$ and $(F1-\bar{F1})||F6$ systems, however, lies
in the fact that as the brane and the antibrane approach other, i.e., as $a\rightarrow 0$, 
they do merge and hence annihilate consistently with the fact that the introduction of the
$NSNS$ $F6$-brane content plays no role as far as in eliminating the semi-classical instability
of these systems. To see this in an explicit manner, we take the $F1-\bar{F1}$ system, for
example, and take the limit $a\rightarrow 0$. First, in the absence of the $NSNS$ $F6$-brane
content,
\beqa
ds^2_{10} &=& \left(1 - {2m\over r}\right)[-dt^2 + dx^2_{1}] + \sum^{6}_{m=2} dx^2_{m} + dr^2
+ r^2\left(1 - {2m\over r}\right)(d\theta^2 + \sin^2 \theta d\phi^2), \nonumber \\ 
e^{2\phi} &=& \left(1 - {2m\over r}\right), ~~~B_{[2]} = 0  
\eeqa 
where we used $\Sigma \rightarrow r^2$, $\Delta \rightarrow r^2(1-2m/r)$, and hence 
$H \rightarrow (1-2m/r)^{-1}$ as $a\rightarrow 0$. In this limit, it appears that 
the opposite electric $NSNS$ charges
carried by $F1$ and $\bar{F1}$ annihilate each other since $B_{[2]}=0$ and the metric solution
now has the topology of $R\times R^{7}\times S^{2}$. Particularly, the manifest $SO(3)$-isometry in
the transverse space implies that, as they approach, $F1$ and $\bar{F1}$ actually merge
and as a result a curvature singularity develops at the center $r=0$. On the other hand, 
consider the $a\rightarrow 0$ limit of the $(F1-\bar{F1})||F6$ solution
\beqa
ds^2_{10} &=& \left(1 - {2m\over r}\right)[-dt^2 + dx^2_{1}] + \sum^{6}_{m=3} dx^2_{m} + dr^2
\nonumber \\
&+& r^2\left(1 - {2m\over r}\right)[d\theta^2 + \sin^2 \theta (d\tilde{\phi} + Bdy)^2] + dy^2, 
\\
e^{2\phi} &=& \left(1 - {2m\over r}\right), ~~~B_{[2]} = 0. \nonumber  
\eeqa 
Evidently, even in the presence of non-zero magnetic field, i.e., $B\neq 0$, the transverse
space still exhibits $SO(3)$-isometry. The only effect of the non-zero $NSNS$ magnetic field is
to endow the transverse $(\tilde{\phi}, y)$ sector (where $y=x_{2}$) with {\it non-trivial 
global topology} and the local geometry of the transverse $(\theta, \tilde{\phi})$ sector is 
still that of $S^2$. This indicates that since $NSNS$ $F6$-brane and $F1$ (and $\bar{F1}$) do not
couple directly (since the first is magnetic whereas the second is electrically-charged under
$B_{[2]}$), the $F6$-brane, playing no role in eliminating the semi-classical instability of the
$F1-\bar{F1}$-pair, simply cannot keep them from colliding and annihilating each other when
$F1$ and $\bar{F1}$ are brought close enough together. And it should be clear that essentially
the same is true for the case of $NS5-\bar{NS5}$ system. Namely as $a\rightarrow 0$, $NS5$ and
$\bar{NS5}$ do merge and annihilate each other despite the presence of the $NSNS$ $F6$-brane.
Since this result is consistent with our naive expectation on the semi-classical endpoint of the
``$NS$'' brane-antibrane systems, we should feel comfortable with this conclusion. Nevertheless,
one may be rather bewildered as this natural endpoint of the ``$NS$'' brane-antibrane systems
turned out to be in sharp contrast with the puzzling picture of the semi-classical endpoint of
the ``$R$'' brane-antibrane systems in the presence of the fluxbrane we discussed earlier.
There, we observed that, despite its failure to serve to stabilize the $D_{2p}-\bar{D}_{2p}$ 
systems, the $RR$ $F7$-brane played the role of keeping them from annihilating each other
completely when $D_{2p}$ and $\bar{D}_{2p}$ are brought close enough together. And we attributed
this apparent puzzle to the limitation of the semi-classical supergravity description which, for
very small inter-brane separation of order the string scale, has to be replaced by the stringy
description in which the $F7$-brane ``delays'' the ``$R$'' brane-antibrane annihilation process
but only until the $F7$-brane itself decays. \\
Of course, there should be a resolution to this contrasting natures between the ``$NS$''-charged
case and ``$R$''-charged case and it appears to be due to the fact that in the ``$NS$''-charged 
case, there is simply no corresponding stringy description of the instability for very small
separations. To be more precise, unlike in the ``$R$''-charged case in which fundamental string
($F1$) ending on $D_{p}$ (and $\bar{D}_{p}$) represented by the brane intersection rule,
$(0|D_{p}, F1)$, develops, in its spectrum, a tachyonic mode which condenses as 
$D_{p}$ and $\bar{D}_{p}$ annihilate, in the ``$NS$''-charged case, the fundamental string does
not end on another fundamental string nor on $NS5$-brane, i.e., no intersection rules such as
$(0|F1, F1)$ or $(0|F1, NS5)$ exists. The only possibilities known for the intersections among
the $NS$-branes in IIA/IIB theories (deduced from $T$ and $S$ duality transformations) 
are \cite{intersect} ;
$(1|F1, NS5)$, $(3|NS5, NS5)$. Thus there are simply no fundamental open strings connecting
$F1-\bar{F1}$ or $NS5-\bar{NS5}$-pair and hence no associated tachyonic modes that replace the
semi-classical instability of these $NS$ brane-antibrane systems for very small separations to
begin with. And on the side of the $NSNS$ $F6$-brane (constructed via the KK-reduction as has
been discussed earlier), it carries no magnetic 3-form flux $H_{[3]}$ to potentially shift the
spectrum of open strings, if any. To summarize, in the ``$NS$''-charged case, the stringy 
description of the instability is simply absent and only the semi-classical supergravity one
exists and according to it, regardless of the presence or absence of the $NSNS$ $F6$-brane,
$F1-\bar{F1}$ and $NS5-\bar{NS5}$ systems are destined to collide and annihilate. And we only
conjecture that the end points would be supersymmetric vacua. \\
Thus far, we have argued, in the ``$NS$''-charged case, that the (open) stringy description is
absent to represent the quantum instability in the $(NS5-\bar{NS5})||F6$ and $(F1-\bar{F1})||F6$ 
systems. This interpretation may provide a resolution to the contrasting features
between the instability of ``$R$''-charged brane-antibrane systems and that of `$NS$''-charged
ones. Certainly, however, it poses another puzzle that in the ``$NS$''-charged case, the quantum
entity, that should take over the semi-classical instability as the inter-brane distance gets
smaller, is missing. Although this is rather an embarrassing state of affair, there indeed appears 
to be an way out as long as the $NSNS$ $F6$-brane content is present in the ``$NS$''-charged 
brane-antibrane systems. To get right to the point,
in the ``$NS$''-charged case the tachyonic modes, that the closed string sector in the $NSNS$
$F6$-brane background develops, appears to be responsible for the quantum instabilities
of the $F1-\bar{F1}$ and $NS5-\bar{NS5}$ pairs as well as for that of the $NSNS$ $F6$-brane itself.
The rationale for this argument has its basis on the work of Russo and Tseytlin \cite{russo1, russo2} 
in which
they demonstrated in an explicit manner that the closed string in the background of the $NSNS$
$F6$-brane given in eq.(42) (which is a special case of the more general species of the $NSNS$
$F6$-brane) develops a tachyonic mode provided the magnetic field strength parameter $B$ is
greater than the critical value, $B> B_{cr}={(R/2\alpha')}$ or equivalently if the radius of the
M-theory circle is smaller than some critical value, $R\leq R_{cr}=\sqrt{2\alpha'}$. And this
happens when no oscillator degrees are excited and for zero KK-momentum mode but when there is a
lowest winding mode along the M-theory circle. And presumably such closed string tachyonic modes
may survive even when the ``$NS$''-charged brane-antibrne pairs are present as well. \\
Thus to summarize, in the ``$R$''-charged case, both the open string tachyonic mode living in the 
$D_{2p}-\bar{D}_{2p}$ systems and the closed string tachyonic mode presumably associated
with the non-supersymmetric and hence unstable $RR$ $F7$-brane are expected to contribute
to the decays of both the fluxbrane and the brane-antibrane systems. Meanwhile in the 
``$NS$''-charged case, the closed string tachyonic mode alone known to arise due to the unstable 
$NSNS$ $F6$-brane with large magnetic field strength (as has been described above) appears to
be responsible for the quantum instabilities and hence the decay of both the fluxbrane and
the brane-antibrane systems presumably via some mechanism such as the condensation. 
This suggested resolution, however, is still not without limitation.
Namely, one might wonder what happens if one erases the $NSNS$ $F6$-brane contents in the
$(NS5-\bar{NS5})||F6$ and $(F1-\bar{F1})||F6$ systems. Even then, will this picture still
holds true ? That is, might the closed strings living in the bulk play some role regarding the quantum
instability of the $F1-\bar{F1}$ and $NS5-\bar{NS5}$ pairs as well ?  At the present stage of
the development of the physics of unstable brane systems, this question cannot be answered in any
definite fashion yet but certainly needs to be considered in a serious manner.

\section{Summary and discussions}\label{ }

In the present work, we raised and then resolved all the relevant puzzles concerning at least 
the semi-classical instabilities of the ``$R$''-charged and ``$NS$''-charged brane-antibrane
systems in type IIA-theory. And in order to intersect $D0-\bar{D0}$ pair with the $RR$ $F7$-brane and
$NS5-\bar{NS5}$ and $F1-\bar{F1}$ pairs with the $NSNS$ $F6$-brane, we had to, along the way,
uplift these solutions to $D=11$ using the M/IIA duality. In this way, we have constructed
$W-\bar{W}$ (i.e., M-wave/anti-M-wave) in eq.(26), $M5-\bar{M5}$ in eq.(34) and $M2-\bar{M2}$
in eq.(46), respectively and to our knowledge, these supergravity solutions representing the M-theory 
brane-antibrane systems have not been discussed in the literature yet and hence make their first
appearance in the present work. Next related to this, it is our next curiosity what the relevant
avenue would be toward the study of instability of ``$R$''-charged brane-antibrane systems in
type IIB-theory such as $D1-\bar{D1}$ and $D5-\bar{D5}$ systems of which the explicit
supergravity solutions are known \cite{youm} (it is rather curious that the $D3-\bar{D3}$ solution is not
known \cite{youm} nor can be obtained via the $T$-dual transformations from the known $D1-\bar{D1}$ 
or $D5-\bar{D5}$ solution). 
Also it seems worthy of note that the results of the analysis presented in this work suggest
that the semi-classical description, based on the supergravity solutions, for the instabilities 
of the ``$R$''-charged brane-antibrane systems is indeed consistent with the stringy description 
in terms of Sen's argument on the endpoint of the unstable branes. We now elaborate on this point.
Firstly, in the absence of the magnetic $RR$ $F7$-brane, the behavior of the supergravity solutions
representing $D_{2p}-\bar{D}_{2p}$ systems for $a\rightarrow 0$ exhibits that as they approach each
other, the brane and the antibrane actually merge and develop curvature singularity at the center,
$r=0$. In the presence of the $RR$ $F7$-brane, however, the behavior of the corresponding
supergravity solutions for $a\rightarrow 0$ indicates that the $RR$ $F7$-brane content of the 
solution plays the role of keeping the brane and the antibrane from annihilating each other 
completely since the two-brane configuration structure still persists in the supergravity solution
even for for very small separation. And in terms of the stringy description, we interpreted this as 
representing that the $RR$ $F7$-brane ``delays'' the brane-antibrane annihilation process by 
introducing an additional energy density to the total energy density or equivalently by providing 
a non-supersymmetric background that survives all the way but only until this non-supersymmetric and 
hence unstable $F7$-brane itself decays. Obviously, this phenomenon of the delay of brane-antibrane 
annihilation by the $RR$ $F7$-brane is a generic stringy effect depending crucially on the tachyonic
mode in the string spectrum and the supersymmetry argument. Thus one would not expect it to have 
any non-supersymmetric point particle field theory analog. Nevertheless, it is amusing to realize that
at least this effect is not counter-intuitive when compared with its counterpart in ordinary point
particle field theory. That is, consider the (electrically) charged particle-antiparticle annihilation
in the presence of a strong external electric field. Were it not for the external field, nothing
could stop the particle-antiparticle pairs from annihilating each other. The external electric field,
however, would relax its strength via the Schwinger process of particle pair creation until it exhausts
all of its energy. Due to this continuous creation of particle-antiparticle pairs while the external
field is alive, the over-all pair annihilation process would slow down, namely, an effective delay of
the pair annihilation would take place. Once the external field vanishes by converting all of its
energy into the particle pair creations, then the usual pair annihilation in the free space will
resume. Again, although this example is not a relevant analog of our brany process, this comparison 
appears  to indicate that the brany phenomenon we discussed above 
does not look so unphysical after all. \\ 
Lastly, we have demonstrated that the behavior of the supergravity solutions representing  
$F1-\bar{F1}$ and $NS5-\bar{NS5}$ for $a\rightarrow 0$ reveals that as they approach, these
``$NS$''-charged brane and antibrane always collide and annihilate irrespective of the presence 
or the absence of the $NSNS$ $F6$-brane. And we have essentially attributed this to the absence 
of (open)
stringy description of the instability in the ``$NS$''-charged case, namely fundamental open string
does not end on another fundamental string nor on $NS5$-brane. We find that all these results from
the semi-classical analysis based on explicit supergravity solutions serve as indirect evidences
supporting Sen's argument for the evolution of unstable $D_{p}-\bar{D}_{p}$ system according to 
which as the separation between the pair becomes of order the string scale, the open string
connecting $D_{p}$ and $\bar{D}_{p}$ develops a tachyonic mode and the $D_{p}-\bar{D}_{p}$ pair
annihilates to a supersymmetric vacuum as the associated open string tachyon condenses, i.e.,
rolls down to a minimum of its potential. \\
We now would like to add more words in this direction. As has been demonstrated in the present
work, from the semi-classical perspective based on relevant supergravity solutions, the 
endpoint of unstable $D_{p}-\bar{D}_{p}$ system is represented by merging and subsequent ``collapse''
of the brane and the antibrane. And according to the $a\rightarrow 0$ limit of the supergravity
solutions representing $D_{p}-\bar{D}_{p}$ systems, the outcome of this collapse turns out to be
a neutral black $p$-brane (since the opposite $RR$ charges are cancelled) having a ``singular''
horizon at $r=2m$ (with $m$ being the brane tension) as well as the curvature singularity at the
center $r=0$. Meanwhile as has been suggested by Sen, from the stringy perspective based on the
open string field theory, the eventual fate of the non-BPS $D_{p}-\bar{D}_{p}$ system could be a
supersymmetric vacuum via the open string tachyon condensation. Namely, the brane and the antibrane
merge and annihilate each other completely since firstly, the opposite $RR$ charges are cancelled
and secondly, the total energy of the system, upon merging, may vanish \cite{sen1}
\beq
E_{tot} = V(T_{0}) + 2M_{D} = 0. 
\eeq  
Thus according to this conjecture by Sen, the outcome of the
brane-antibrane collision could be a complete annihilation into a supersymmetric vacuum.
In the ``$NS$''-charged case, however, the situation changes as we have seen in the text. There, in
terms of the semi-classical description based on the exact supergravity solutions, the endpoint of
unstable $F1-\bar{F1}$ or $NS5-\bar{NS5}$ system still appears to be merging and the ``collapse''.
And hence the outcome of this collapse is again the neutral black string or black 5-brane having
singular horizon at $r=2m$ as well as the curvature singularity at the center $r=0$. However, since
fundamental string does not end on another $F1$ nor on $NS5$ (namely no intersection rules such 
as $(0|F1, F1)$ or $(0|F1, NS5)$ exists), there is as a result, no stringy description available 
for the brane-antibrane annihilation in terms of open string tachyon condensation via Sen's mechanism.
This absence of the quantum mechanism for the outcome of $F1-\bar{F1}$ or $NS5-\bar{NS5}$ annihilation
is indeed a very unnatural state of affair in light of the fact that 
$F1-\bar{F1}$ and $NS5-\bar{NS5}$ systems are just $U=S T$ duals to $D2-\bar{D2}$ and to
$D6-\bar{D6}$ systems respectively. Certainly, therefore, a quantum, stringy description is in need
for these $F1-\bar{F1}$ and $NS5-\bar{NS5}$ annihilations into (presumably) supersymmetric vacua.

\section*{Acknowledgments}

This work was supported in part by grant No. R01-1999-00020 from the Korea Science and Engineerinig 
Foundation and by BK21 project in physics department at Hanyang Univ.

\appendix

{\rm \bf \large Appendix}

\section{Intersecting $D0-\bar{D0}$ with an electric $RR$ fluxbrane}

As we mentioned earlier, since $D0-\bar{D0}$ and $D2-\bar{D2}$ systems are ``electrically''
$RR$ charged, it may be natural to attempt to intersect them with electric $RR$ fluxbranes.
Attempt of this sort, however, via the twisted KK-reduction of $W-\bar{W}$ and $M2-\bar{M2}$
systems respectively, fails since the electric $RR$ fluxbranes constructed in this
way turns out to be essentially trivial. Thus in this appendix, by taking the $D0-\bar{D0}$
case for example, we shall show in an explicit manner that this is what actually happens. \\
Start again with the $W-\bar{W}$ solution in $D=11$ M-theory given by eq.(26) in the text.
We first write the transverse coordinates as $(y=x_{1}, x^{m}, r, \theta, \phi)$ $(m=1,..,6)$
with $y$ being identified with the coordinate on the M-theory circle. Next, we consider the
{\it twisted} KK-reduction of this D=11 solution. However, since we are interested in generating
an ``electrically'' $RR$-charged fluxbrane, we now choose to perform the following non-trivial
point identification
\beq
(y, t) \equiv (y+2\pi n_{1}R, t+2\pi n_{1}R^{2}E), ~~~n_{1}\in Z
\eeq
(with $E$ being the electric field parameter) followed by the associated skew KK-compactification 
along the orbit of the Killing field
\beq
l = \left(\partial/\partial y\right) + ER \left(\partial/\partial t\right).
\eeq
In the twisted KK-reduction of this type, however, one should worry about the emergence of 
closed timelike curves. Therefore, we introduce the {\it adapted} coordinate
\beq
\tilde{t} = t - ERy
\eeq
which is constant along the orbits of $l$ and in terms of which, the metric is free of
closed timelike curves since now the adapted time coordinate $\tilde{t}$ has standard 
semi-infinite range $0 \leq \tilde{t} <\infty$. Finally, we proceed with the standard
KK-reduction along the orbit of $(\partial/\partial y)$. The result is
\beqa
ds^2_{10} &=& \tilde{\Lambda}^{1/2}\left\{-H^{-1/2}d\tilde{t}^2 + H^{1/2}[\sum^{6}_{m=1}dx^2_{m} +
(\Delta + a^2\sin^2 \theta)\left({dr^2\over \Delta} + d\theta^2\right) + 
\Delta \sin^2 \theta d\phi^2]\right\} \nonumber \\
&+& \tilde{\Lambda}^{1/2}H^{3/2}\left\{A^2_{t} - \tilde{\Lambda}^{-1}[A_{t}(1 + A_{t}ER)
- H^{-2}ER]^2\right\}d\tilde{t}^2, \nonumber \\
e^{{4\over 3}\phi} &=& H\tilde{\Lambda}, \\
A_{[1]} &=& \tilde{\Lambda}^{-1}[A_{t}(1 + A_{t}ER)-H^{-2}ER]d\tilde{t}, \nonumber \\
F_{[2]} &=& (\partial_{r}A_{\tilde{t}})dr\wedge d\tilde{t} +
(\partial_{\theta}A_{\tilde{t}})d\theta\wedge d\tilde{t},
~~~{\rm where} \nonumber \\
A_{t} &=& {2ma\cos \theta \over \Sigma},
~~~\tilde{\Lambda} = [(1+A_{t}ER)^2 - H^{-2}E^{2}R^{2}]. \nonumber
\eeqa
We next consider the conical singularity structure of this new solution.
To do so, notice that again in this case,
$\psi^{\mu}\psi_{\mu}=g_{\tilde{\phi}\tilde{\phi}}=0$ still has roots at
the locus of $r=r_{+}$ as well as along the semi-infinite axes
$\theta =0, \pi$. Thus we need to worry about the possible
occurrence of conical singularities both along $\theta=0, \pi$ and
at $r=r_{+}$ again. Assuming that the azimuthal angle coordinate
$\tilde{\phi}$ is identified with period $\Delta \tilde{\phi}$,
the conical deficit along the axes $\theta =0, \pi$ and along the
segment $r=r_{+}$ are given respectively by
\beqa
\delta_{(0,\pi)} &=& 2\pi - \arrowvert {\Delta \phi
d\sqrt{g_{\phi\phi}} \over \sqrt{g_{\theta\theta}}d\theta}
\arrowvert_{\theta =0, \pi} = 2\pi - \Delta \phi, \\
\delta_{(r=r_{+})} &=& 2\pi - \arrowvert {\Delta \phi
d\sqrt{g_{\phi\phi}} \over \sqrt{g_{rr}}dr} \arrowvert_{r=r_{+}} =
2\pi - \left(1 + {m^2\over a^2}\right)^{1/2}\Delta \phi \nonumber
\eeqa 
where, of course, we used the metric solution given in eq.(61).
Note that the conical singularity structure remains essentially the 
same as those for the $D0-\bar{D0}$ system despite the introduction
into the system an electric $RR$ fluxbrane content. Indeed, this has
been expected since the electric fluxbrane content we attempted to 
introduce via this twisted KK-reduction turns out to be a trivial one
having null structure. In order to get the physical explanation for this
failure, we examine the nature of the electric fluxbrane content. The 
pure electric fluxbrane content in this $(D0-\bar{D0})||({\rm electric ~fluxbrane})$
solution given in eq.(61) can be extracted in an unambiguous manner simply by 
erasing the brane-antibrane content. That is, by setting $m=0$ and $a=0$,
we get
\beqa
ds^2_{10} &=& \tilde{\Lambda}^{1/2}[-d\tilde{t}^2 + \sum^{6}_{i=1}dx^2_{i}
+ dr^2 + r^2 (d\theta^2 + \sin^2 \theta d\phi^2)] - \tilde{\Lambda}^{-1/2}
E^2 R^2 d\tilde{t}^2 \nonumber \\
&=& \tilde{\Lambda}^{1/2}[-dt^2 + \sum^{6}_{i=1}dx^2_{i} + dr^2 + 
r^2 (d\theta^2 + \sin^2 \theta d\phi^2)], \\
F_{[2]} &=& 0 ~~~({\rm vacuum}) \nonumber 
\eeqa
where $\tilde{\Lambda} = (1 - E^2 R^2)$ and in the last line, we redefined
$t = (1+\tilde{\Lambda}^{-1}E^2 R^2)^{1/2}\tilde{t}$. This is essentially a
flat spacetime. And it means that although we attempted to introduce into
the $D0-\bar{D0}$ system an electric fluxbrane via the twisted KK-reduction
by mixing the orbits of the Killing fields $(\partial/\partial y)$ and
$(\partial/\partial t)$, it turns out that no non-trivial electric fluxbrane
was generated. Precisely due to this null nature of the electric fluxbrane, 
the conical singularity structure remained unchanged as we have seen above.
Lastly, it seems noteworthy that for the earlier twisted KK-reduction in which
the orbits of the Killing fields $(\partial/\partial y)$ and 
$(\partial/\partial \phi)$ were mixed, the non-trivial magnetic fluxbrane was
generated whereas for the present case when those of the Killing fields
$(\partial/\partial y)$ and $(\partial/\partial t)$ are mixed, trivial fluxbrane
with null structure results. It is interesting to note that indeed, this is 
reminiscent of an well-known solution-generating technique in 4-dimensional
Einstein-Maxwell theory \cite{wald} which can be stated as : {\it the axial Killing vector
$\psi^{\mu}=(\partial/\partial \phi)^{\mu}$ in a vacuum spacetime generates a
stationary, axisymmetric test electromagnetic field which asymptotically
approaches a uniform magnetic field whereas the time-translational Killing
vector $\xi^{\mu}=(\partial/\partial t)^{\mu}$ in a vacuum spacetime generates
a stationary, axisymmetric electromagnetic field which vanishes asymptotically.}

\end{document}